\documentclass[letterpaper,twocolumn,prb,aps,showpacs,floatfix, 
superscriptaddress]{revtex4-2}
\usepackage[latin1]{inputenc}
\usepackage{bm}
\usepackage[usenames]{color}
\usepackage{multirow}
\usepackage{amssymb}
\usepackage{amsbsy}
\usepackage{amsmath}
\usepackage{stmaryrd}
\usepackage{graphicx}
\usepackage{epsfig}
\usepackage{placeins}
\usepackage{bbold}
\usepackage{bbm}
\usepackage{physics}
\usepackage[colorlinks,linkcolor=blue,citecolor=blue,urlcolor=blue]{hyperref}

\usepackage{scalerel}
\usepackage{tikz}
\usetikzlibrary{calc}
\usetikzlibrary{patterns}
\usetikzlibrary{svg.path}
\definecolor{orcidlogocol}{HTML}{A6CE39}
\tikzset{
  orcidlogo/.pic={
    \fill[orcidlogocol] 
svg{M256,128c0,70.7-57.3,128-128,128C57.3,256,0,198.7,0,128C0,57.3,57.3,0,128,
0C198.7,0,256,57.3,256,128z};
    \fill[white] svg{M86.3,186.2H70.9V79.1h15.4v48.4V186.2z}
                 
svg{M108.9,79.1h41.6c39.6,0,57,28.3,57,53.6c0,27.5-21.5,53.6-56.8,
53.6h-41.8V79.1z 
M124.3,172.4h24.5c34.9,0,42.9-26.5,
42.9-39.7c0-21.5-13.7-39.7-43.7-39.7h-23.7V172.4z}
                 
svg{M88.7,56.8c0,5.5-4.5,10.1-10.1,10.1c-5.6,0-10.1-4.6-10.1-10.1c0-5.6,4.5-10.1
,10.1-10.1C84.2,46.7,88.7,51.3,88.7,56.8z};
  }
}

\newcommand\orcid[1]{\!%
  \href{https://orcid.org/#1}{%
    \mbox{%
      \scaleto{%
        \begin{tikzpicture}[yscale=-1,transform shape]
          \pic{orcidlogo};
        \end{tikzpicture}
      }{8pt}%
    }%
  }%
}

\makeatletter

\begin{document}
\title{Temporal relaxation of disordered 
many-body quantum systems under 
driving and dissipation}

\author{Jonas Richter~\orcid{0000-0003-2184-5275}}
\affiliation{Department of Physics, Stanford University, Stanford, CA 94305, 
USA}
\affiliation{Institut f\"ur Theoretische Physik, Leibniz 
Universit\"at Hannover, 30167 Hannover, Germany}

\date{\today}

\begin{abstract}
Strong disorder inhibits thermalization 
in isolated quantum systems and may lead to 
many-body localization (MBL). In realistic situations, however,  
the observation of MBL is hindered by residual couplings of 
the system to an environment, which acts as a bath and pushes the system to thermal equilibrium. This paper is concerned 
with the transient dynamics prior to thermalization and studies how the 
relaxation of a disordered system is altered under the 
influence of external 
driving and dissipation. We consider a scenario where 
a disordered quantum spin chain 
is placed into a strong magnetic field that 
polarizes the system. By suddenly removing the 
external field, 
a nonequilibrium situation is induced and the decay of 
magnetization probes the degree of localization. We 
show that by driving the system with light, one can distinguish between 
different 
dynamical regimes as the spins 
are more or less susceptible to the drive 
depending on the strength of the disorder. 
We provide evidence that some of these signatures remain observable at 
intermediate time scales 
even when the 
spin chain 
is subject to noise due to 
coupling to an environment. From a numerical point of view, 
we demonstrate that the open-system dynamics 
starting
from a class of 
experimentally relevant mixed initial states can be 
efficiently simulated by 
combining dynamical quantum typicality 
with stochastic unraveling of Lindblad 
master equations. 
 
\end{abstract}

\maketitle


\section{Introduction}

Generic many-body quantum systems prepared in some out-of-equilibrium initial state
are expected to relax to thermal equilibrium at long times \cite{Eisert_2015, 
Nandkishore_2015}. In strongly 
disordered systems, the process of thermalization can be slowed down and may 
potentially cease entirely due to many-body localization (MBL) 
\cite{Nandkishore_2015, Abanin_2019}. While numerous studies have found evidence 
for MBL in disordered one-dimensional systems (e.g., \cite{Pal_2010, Kj_ll_2014, Luitz_2015, Imbrie_2016}), the asymptotic 
stability of MBL as a nonequilibrium phase of 
matter is still under debate at present \cite{_untajs_2020, Sels_2021, 
Morningstar_2022, Abanin_2021}. The main 
complication stems from the necessity of studying large system sizes 
and long time scales, which is beyond the reach of state-of-the-art 
numerical approaches \cite{Doggen_2018, Panda_2020, Richter_2022}. Similarly, while 
groundbreaking experiments in cold-atom 
or trapped-ion platforms have immensely contributed to our understanding of 
disordered many-body quantum dynamics, they cannot unambiguously 
confirm the 
(non)existence of MBL based on intermediate-time signatures \cite{Choi_2016, 
Schreiber_2015, Smith_2016}. In contrast to
such quantum-simulator platforms, it is even more 
challenging to observe MBL in traditional solid-state experiments as 
the inevitable coupling of the spin or electron system to the phonons provides 
a heat bath that favors thermalization \cite{Ovadia_2015, Nietner2022}.

The phenomenology of MBL is typically understood with respect to local 
integrals of motion, so-called l-bits \cite{Serbyn_2013, Huse_2014, Chandran_2015, 
Ros_2015}. Due to overlap with these l-bits, local 
observables fail to relax to thermal equilibrium under time evolution. 
When 
coupled to a thermal bath \cite{Fischer_2016, 
Levi_2016, Medvedyeva_2016, Everest_2017, Gopalakrishnan_2017, Lazarides_2017, 
Wu_2019, Ren_2020}, MBL is typically believed to be unstable 
\cite{Nandkishore_2016, De_Roeck_2017, Luitz_2017}, 
as also observed in cold-atom experiments \cite{L_schen_2017, Rubio_Abadal_2019}. 
However, interesting dynamical regimes might emerge if the coupling is weak or 
the bath is small \cite{Huse_2015, Marino_2018}. 
Moreover, driving a disordered system has been shown to enable the realization of novel concept such as time crystals 
\cite{Khemani_2019, Zalatel2023}.      
Generally, the combination of a bath and an external drive opens up a vast landscape of driven-dissipative systems with exotic out-of-equilibrium phenomena \cite{Sieberer2023}.

\begin{figure}[tb]
 \centering
 \includegraphics[width=0.95\columnwidth]{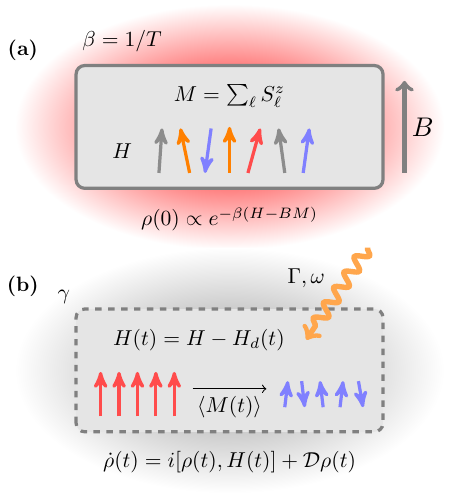}
 \caption{Sketch of the setup. {\bf (a)} Initial-state preparation. A 
disordered quantum spin chain 
 described by a Hamiltonian $H$ is subject to a (strong) magnetic field $B$ 
that couples to the global magnetization $M$. The system 
is weakly connected to a heat 
bath at inverse temperature $\beta$, and 
thermalization to 
a Gibbs state 
$\rho(0) \propto \exp[-\beta(H-BM)]$ is assumed. 
{\bf (b)} Dynamics. At $t = 0$, the 
magnetic field is suddenly switched off, resulting in a nonequilibrium 
situation (i.e., a quantum quench), and the magnetization $\langle 
M(t)\rangle$ will decay in time depending on the disorder strength $W$. 
We study how $\langle M(t) \rangle$ is altered 
by driving the system with circularly polarized light 
with amplitude $\Gamma$ and frequency $\omega$, as well as by 
considering dissipation modelled by a Lindblad master equation with 
system-bath coupling $\gamma$. }
 \label{Fig:Sketch}
\end{figure}

In this paper, we study how the dynamics 
of a disordered system changes 
when subjected to certain driving protocols and a noisy environment.
We do not aim to resolve the question 
of whether MBL asymptotically exists 
as a stable phase of matter, neither in open nor 
in closed systems. Rather, we ask more generally if certain features of the dynamics of disordered 
quantum systems leave a fingerprint in a potentially experimentally 
feasible setup. To this end, we particularly build on an idea  
proposed by Ros and M\"uller \cite{Ros_2017}, who considered the remanent 
magnetization 
of an antiferromagnetic spin chain that is initially polarized in a 
ferromagnetic state. Specifically, consider a spin system, 
which is fully polarized 
by a strong magnetic field, cf.\ Fig.\ \ref{Fig:Sketch}~(a). 
At some point, the field 
is switched off and the 
magnetization will decay towards 
a long-time value which is nonzero in the case 
of MBL. In contrast to typical probes of MBL that require highly 
local resolution, the total magnetization has 
the advantage that it should in 
principle be easier 
measurable in solid-state experiments. 
While Ref.\ \cite{Ros_2017} focused on the ideal 
situation of unitary time evolution, we 
here go beyond these results 
and consider a scenario where the decay of the 
magnetization is altered due to the influence of an
environment, cf.\ Fig.\ \ref{Fig:Sketch}~(b). 

While we expect the system 
to relax to thermal equilibrium
when coupled to a thermal bath, 
another motivation for our work 
stems from studies by Lenar\v{c}i\v{c} {\it et al.}, who argued 
that certain 
features of MBL can be reactivated by driving the system 
with light \cite{Lenar_i__2018, Lenar_i__2020}. In contrast to
Ref.\ \cite{Lenar_i__2018}, which explored the emergence of characteristic features in the steady state, we here study 
the possibility of 
using the drive to induce observable signatures in the transient dynamics on intermediate time scales. In this context, we follow earlier works, which showed that 
circularly polarized light can be used to induce a magnetization in 
strongly anisotropic spin chains \cite{Takayoshi_2014, Takayoshi_2014_2, 
Herbrych_2016}. 
Combined with our setup, 
we demonstrate that this type of driving can be used to alter the temporal relaxation in disordered 
spin chains. 

Summarizing our main results, we show that the presented protocol allows to distinguish between systems with weak disorder (which show a strong response to the drive) and systems with strong disorder (which show a weaker response to the drive). In particular, for suitable values of driving amplitude and frequency, we find that the relaxation of weakly disordered systems is significantly slowed down due to the drive-induced nonequilibrium magnetization. In contrast, at stronger disorder, the drive appears to facilitate the relaxation of magnetization and weakens the system's tendency to localize.  

While our focus is on fully polarized pure initial states, we also consider mixed states at finite temperatures and 
different initial polarizations, i.e., states that are prepared close to as well as 
far away from equilibrium. 
In this context, from a numerical point of view, we show that 
the dynamics resulting from 
this class of nonequilibrium 
states can be efficiently simulated by exploiting the 
typicality of random pure quantum states. In particular, 
we demonstrate that dynamical quantum typicality (DQT)
provides a useful approach even if the system is 
coupled to an environment by combining 
DQT with stochastic unraveling 
of Lindblad master 
equations. This numerical combination to study open-system dynamics 
might be of independent interest also in other 
contexts due to its 
independence on details of the system, 
the environment, and the driving 
protocol. 

The rest of this paper is structured as follows. 
In Sec.\ \ref{Sec::Setup}, we define the models and observables studied in this 
work. Our numerical approach is discussed in 
Sec.\ \ref{Sec::Numerics}, and we present our results in Sec.\ 
\ref{Sec::Results}. We summarize and conclude in Sec.\ 
\ref{Sec::Conclu}.


\section{Setup}\label{Sec::Setup}

We consider a spin chain with $L$ sites and periodic 
boundary 
conditions, 
\begin{equation}\label{Eq::Ham0}
 H = \sum_{\ell = 1}^L \sum_{\mu = x,y,z} J^\mu S_\ell^\mu S_{\ell+1}^\mu + 
\sum_{\ell = 1}^L h_\ell S_\ell^z\ , 
\end{equation}
where $S_\ell^\mu = \tfrac{1}{2} \sigma_\ell^\mu$ are spin-$1/2$ operators, and 
the on-site fields $h_\ell$ are drawn at random from a uniform 
distribution, $h_\ell \in [-W,W]$, with $W$ setting the disorder strength. For 
$J^{x,y,z} = 1$, Eq.\ \eqref{Eq::Ham0} reduces to the disordered 
Heisenberg chain which is well studied in the 
context of MBL. In this paper, however, we are particularly interested in the case of 
anisotropic couplings with nonconserved global magnetization $M$, i.e., 
\begin{equation}
 M = \sum_{\ell = 1}^L S_\ell^z\ ,\quad [H,M] \neq 0\ .
\end{equation}
For concreteness, we will set $J^x = J^z = 1$ and $J^y = 0$ \cite{Ros_2017},
but we expect that our findings will qualitatively carry over also 
to other choices of the exchange couplings. 
Moreover, in Appendix \ref{Sec::App::Heis} we present additional 
results for the standard MBL model with $J^x = J^y = J^z = 1$.

We consider a nonequilibrium protocol as in 
\cite{Ros_2017}, 
where the spin system is initially placed in a strong 
external magnetic field. 
Let us further assume that the system is weakly 
coupled to a heat 
bath at inverse temperature $\beta$, such that the situation can be described by a thermal Gibbs state of the 
form
\begin{equation}\label{Eq::Mixed}
 \rho(0) = \frac{e^{-\beta(H-BM)}}{\text{tr}[e^{-\beta(H-BM)}]}\ , 
\end{equation}
where the field of strength $B$ couples to the magnetization operator, see Fig.\ \ref{Fig:Sketch}~(a). At time $t = 0$, the magnetic 
field is suddenly switched off ($B\to0$) and the state $\rho(0)$ is no 
equilibrium state 
of the remaining Hamiltonian $H$, see Fig.\ \ref{Fig:Sketch}~(b). Such 
types of quench protocols involving the sudden removal of an external force have
been studied also in \cite{Richter_2018, Richter_2019, Richter_2019_2}. 

In the limit of strong $B$, the spin 
chain is
fully 
polarized such that $\rho(0) \to \ketbra{\psi}{\psi}$ is 
a pure state 
with $\ket{\psi} =  \ket{\uparrow \uparrow \cdots \uparrow}$. While our focus 
will be on this fully polarized case, we also consider weaker magnetic fields 
where $\rho(0)$ remains a mixed state and one can explore the ensuing dynamics depending on $\beta$ and $B$.  

Given $\rho(0)$, we study the temporal relaxation of the magnetization, 
\begin{equation}
 \langle M(t) \rangle = \text{tr}[\rho(t) M]\ ,
\end{equation}
where we consider different scenarios for the time evolution of $\rho(t)$, namely (i) the isolated system without external driving or dissipation, (ii) the driven system with unitary dynamics, and (iii) the driven-dissipative situation where the system is coupled to an environment. 

In the isolated case, the time evolution is 
understood with respect to $H$ in Eq.\ \eqref{Eq::Ham0}, $\rho(t) = 
e^{-iHt}\rho(0) e^{iHt}$. 
For weak disorder, we expect $\langle M(t) \rangle$ to decay 
towards zero indicating thermalization. In contrast, 
for stronger disorder, the 
decay 
of $\langle M(t) \rangle$ will be slower and the remanent magnetization will remain nonzero on rather long time scales \cite{Ros_2017}, indicating a transition to a
finite-size MBL regime. While we are interested in the behavior of $\langle M(t) \rangle$ on realistic time scales, we do not aim 
to draw conclusions on the asymptotic stability of MBL in the thermodynamic limit.

We also study the possibility 
of altering the dynamics of $\langle M(t) \rangle$ by driving the system. 
Specifically, we here follow the setup in \cite{Takayoshi_2014, Takayoshi_2014_2, 
Herbrych_2016}, which considered 
the induced nonequilibrium magnetization by
circularly 
polarized light propagating in the $z$-direction. Assuming that only the 
magnetic component of the light couples to the system, the resulting 
time-dependent Hamiltonian takes the form,
\begin{equation}\label{Eq::Hdriven}
H(t) = H - H_d(t)\ ,  
\end{equation}
where the driving term $H_d(t)$ is given by
\begin{equation}\label{Eq::DriveDrive}
 H_d(t) = \Gamma \sum_{\ell = 1}^L \left(e^{-i\omega t} S_\ell^+ + e^{i\omega 
t} S_\ell^-  \right)\ . 
\end{equation}
Here, $\Gamma > 0$ is the amplitude and $\omega > 0$ is the frequency of the 
light. In an actual experiment, $\Gamma$ and $\omega$ might be 
time-dependent, which is neglected here. 

Eventually, we want to consider a situation where the 
system is not isolated but actually coupled to an environment. 
For simplicity, we do not attempt to 
describe a concrete microscopic situation, e.g., modeling explicitly 
a phonon bath that would be 
present in a spin-chain material. 
Instead, we treat the environment in terms of a Lindblad master 
equation \cite{Breuer_2007},
\begin{equation}\label{Eq:Lindblad}
\dot{\rho}(t) = {\cal L} \, \rho(t)  = i [\rho(t),H(t)] + {\cal D} \, \rho(t) 
\, ,
\end{equation}
consisting of a unitary time evolution with respect to the (driven) $H(t)$, and 
a dissipative part that is given by,
\begin{equation}\label{Eq:Lindblad2}
{\cal D} \, \rho(t) = \sum_j \gamma \Big ( L_j \rho(t) L^\dagger_j - 
\frac{1}{2} \{ \rho(t), L_j^\dagger L_j \} \Big )\ , 
\end{equation}
where the $L_j$ denote a set of Lindblad jump operators, $\gamma$ is the 
strength of the system-bath coupling, and $\lbrace \cdot,\cdot \rbrace$ denotes 
the anticommutator. Specifically, we consider structureless bulk noise in 
the form of 
dephasing with jump operators at each lattice site, 
\begin{equation}
 L_j = \sigma_j^z\ ,\quad j = 1,\dots,L\ . 
\end{equation}
The 
effect of dephasing noise on the stability of MBL and on transport 
properties in disordered quantum systems has been studied 
both in the Markovian and non-Markovian regime \cite{Fischer_2016, 
Levi_2016, Medvedyeva_2016, Everest_2017, Gopalakrishnan_2017, _nidari__2016, 
Ren_2020}. 
Even though this choice of jump operators conserves the magnetization, 
we will show below that it still facilitates the relaxation of $\langle 
M(t)\rangle$. 

\section{Numerical approach}\label{Sec::Numerics}

In order to study $\langle M(t)\rangle$, we
evolve the out-of-equilibrium initial 
state $\rho(0)$ in time. If $\rho(0) \to 
\ket{\psi} = \ket{\uparrow \uparrow \cdots \uparrow}$ and the system is 
isolated from the environment, we employ standard sparse-matrix 
techniques \cite{Fehske_2009} to  
solve the time-dependent Schr\"odinger equation $\ket{\psi(t)} = 
e^{-iHt}\ket{\psi(0)}$. Since $H(t)$ does not conserve the 
magnetization, these simulations are carried out in the full Hilbert space with 
dimension $2^L$. Moreover, we perform a disorder 
average over approximately $\sim500$ realizations of the 
random on-site fields. As the global observable $\langle M(t)\rangle$ turns out to 
be rather insusceptible to finite-size effects on the time scales considered (see also Appendix \ref{Sec::App_FS}), we restrict ourselves to system sizes $L \approx 18 - 20$ in this paper. While this is already beyond the system sizes accessible to standard ED, we note that even larger systems are in principle amenable to the pure-state techniques used here.   

We can use sparse-matrix 
propagation of pure quantum states also 
in the more general case where $\rho(0)$ is a mixed state. To this end, 
we rely on the concept of dynamical quantum typicality, 
which exploits the properties of random pure quantum states (see \cite{Heitmann_2020, Jin_2021} for reviews).  
Specifically, we 
consider pure states of the form,
\begin{equation}\label{Eq::TypicalState}
 \ket{\psi(0)} = 
\frac{\sqrt{\rho(0)}\ket{\Psi}}{\sqrt{\bra{\Psi}\rho(0)\ket{\Psi}}}\ , 
\end{equation}
where $\ket{\Psi}$ is a random state drawn from the unitarily invariant Haar
measure, i.e., 
\begin{equation}
 \ket{\Psi} = \sum_{k=1}^{2^L} c_k \ket{\varphi_k}\ , 
\end{equation}
where the real and imaginary parts of 
the complex coefficients $c_k$ are drawn 
from a Gaussian distribution with zero 
mean. The application of $\sqrt{\rho(0)}$ on the random state 
$\ket{\psi}$ in Eq.\ \eqref{Eq::TypicalState} can be achieved by an 
imaginary time evolution with respect to $H-BM$ up to $\beta/2$.
Exploiting DQT, the exact dynamics of $\langle 
M(t)\rangle$ can 
be approximated by the expectation value 
within the pure state $\ket{\psi}$,  
\begin{equation}\label{Eq::Typ}
 \langle M(t) \rangle = \text{tr}[\rho(t)M] = \bra{\psi(t)}M\ket{\psi(t)} + 
\epsilon(\ket{\psi})\ , 
\end{equation}
where the statistical error $\epsilon(\ket{\psi})$ scales as 
$1/\sqrt{d_\text{eff}}$, with $d_\text{eff}$ denoting an effective 
Hilbert-space dimension that depends on the choices 
of $\beta$ and $B$ \cite{Hams2000, Sugiura2013}. Crucially, $d_\text{eff}$ grows exponentially with 
increasing system size $L$ such 
that the approximation in Eq.\ \eqref{Eq::Typ}
becomes highly accurate. This accuracy can be 
further improved by 
averaging over different instances of the random state $\ket{\Psi}$. In fact, 
in our simulations, we combine the averaging over states with the 
above-mentioned disorder average by choosing 
a new $\ket{\Psi}$ for each 
random disorder realization. 

Eventually, in order to simulate the dynamics of the system coupled 
to dephasing noise, we rely on 
stochastic unraveling of the Lindblad equation, where 
Eq.\ \eqref{Eq:Lindblad} is approximated by averaging over pure-state 
trajectories that consist of sequences of 
deterministic evolutions and quantum jumps \cite{Dalibard_1992}. Specifically, for each trajectory, 
the 
pure states $\ket{\psi(t)} = e^{-iH_\text{eff}t}\ket{\psi}$ evolve under an 
effective Hamiltonian,
\begin{equation}\label{Eq::Heff}
  H_\text{eff}(t) = H(t) - i \frac{\gamma}{2} \sum_{\ell=1}^L \sigma_\ell^z 
\sigma_\ell^z = H(t) - \frac{i\gamma L}{2}\ .  
\end{equation}
Since $H_\text{eff}$ is non-Hermitian, the norm 
of $\ket{\psi(t)}$ will decrease. Once $||\ket{\psi(t)}|| < \epsilon$ 
drops below a randomly drawn threshold $\epsilon \in [0,1]$, a quantum jump occurs with respect to one of the 
jump operators and the resulting state is normalized, $\ket{\psi(t)} \to 
\ket{\psi'(t)} = \sigma_\ell^z \ket{\psi(t)}/||\sigma_\ell^z 
\ket{\psi(t)}||$. Subsequently, another deterministic evolution with respect to
$H_\text{eff}$ takes place. 

For each trajectory of the stochastic unraveling, the 
initial state is chosen as a  
random realization of $\ket{\Psi}$ in 
Eq.\ \eqref{Eq::TypicalState}. 
Averaging over sufficiently many 
trajectories then approximates the open-system 
dynamics 
of $\langle M(t) \rangle$ with initial state $\rho(0)$, i.e., 
\begin{equation}
 \langle M(t)\rangle \approx \sum_{\bf r}
\frac{\bra{\psi_{{\bf 
r}}(t)}M\ket{\psi_{\bf r}(t)}}{\braket{\psi_{\bf r}(t)}{\psi_{\bf r}(t)}}\ , 
\end{equation}
where the subscript ${\bf r}$ labels trajectories with random quantum jumps. 
Note that the averaging over trajectories can again be performed simultaneously with the 
averaging over typical states and disorder realizations $h_\ell$. 
Using this combination of quantum typicality [to mimic the mixed initial state $\rho(0)$] and stochastic unraveling, 
we here study open systems of sizes $L = 18$, beyond the range of standard exact diagonalization, and with comparable computational costs to simulations of isolated systems discussed above, see also \cite{Heitmann_2022, Heitmann_2023}.


\section{Results}\label{Sec::Results}

We now present our numerical results. 
In Sec.\ 
\ref{Sec::Results_Isolated} we consider the relaxation of the magnetization $\langle 
M(t)\rangle$ in the isolated system $H$. 
These results will provide a useful 
point of reference 
for studying the impact of driving in Sec.\ 
\ref{Sec::Results_Drive}, as well as for the Lindblad dynamics of the open 
system in Sec.\ \ref{Sec::Results_Open}. 

\subsection{Isolated system}\label{Sec::Results_Isolated}

\begin{figure}[tb]
 \centering
 \includegraphics[width = 0.95\columnwidth]{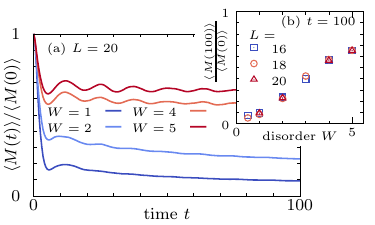}
 \caption{{\bf (a)} Decay of normalized magnetization $\langle M(t) \rangle/\langle 
M(0)\rangle$ 
starting from the 
fully polarized state $\ket{\uparrow}$ for 
$L = 20$ and 
different values of $W$. {\bf (b)} Long-time value $\langle M(t = 100) 
\rangle/\langle M(0)\rangle$ 
versus disorder strength $W$ for different $L$. Data is obtained by averaging 
over approximately $\sim 500$ disorder realizations. }
 \label{Fig1}
\end{figure}

As a starting point, it is instructive to study the dynamics of the isolated 
system in Eq.\ \eqref{Eq::Ham0}. 
For the initial state, we consider a strong  
magnetic field $B$, 
such that $\ket{\psi(0)} = \ket{\uparrow \uparrow \cdots  \uparrow}$. 
In Fig.\ \ref{Fig1}~(a), the relaxation of magnetization $\langle 
M(t) \rangle$ is shown for system size $L = 18$ and different values of the 
disorder $W$, normalized by the initial value 
$\langle M(0) \rangle = L/2$.   
While at weak disorder $\langle M(t) \rangle$ decays towards 
small values 
indicating thermalization,
this decay 
slows down considerably at 
larger 
$W \approx 5$. The remanent 
magnetization can thus be seen as an indicator for 
the strength of disorder and the onset of a MBL regime \cite{Ros_2017}.

We analyze the decay of $\langle M(t) \rangle$ in some more detail 
in Fig.\ \ref{Fig1}~(b) by 
extracting $\langle M(t=100)\rangle$, which is 
found to increase monotonically with $W$.
Moreover, comparing data for different 
systems sizes, we find that at least on the 
time scales $t \leq 100$ shown here, finite-size effects are comparatively 
weak. 

\begin{figure}[tb]
 \centering
 \includegraphics[width=0.95\columnwidth]{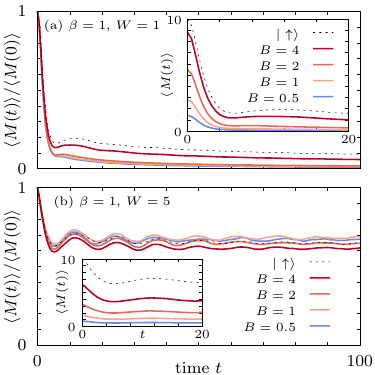}
 \caption{{\bf (a)} $\langle M(t)\rangle$ for initial states $\rho(0)$ at fixed 
inverse temperature $\beta = 1$ and varying magnetic field $B = 0.5,\cdots,4$. 
The dashed curve indicates data for the fully polarized state $\ket{\uparrow}$ 
(i.e., $B \to \infty$). Original dynamics is shown in the inset (shorter times), while the main panel shows the data normalized by the initial value $\langle M(0)\rangle$ and $t\leq 100$. {\bf (b)} Analogous data but now for stronger disorder $W = 
5$. The system size is $L = 20$ in all cases. 
}
 \label{Fig2}
\end{figure}

In addition to the fully polarized initial state, one can 
consider the case of a mixed state $\rho(0)$ that depends 
on the 
inverse temperature $\beta$ and the field strength $B$. 
In particular, it might be conceivable that an actual 
experiment, for example in a solid-state setting, allows at 
least some control over $\beta$ and especially 
$B$ in order to tune $\rho(0)$ into different regimes.
In Figs.\ \ref{Fig2}~(a) and \ref{Fig2}~(b), 
we show $\langle M(t)\rangle$ for $W = 1$ and $W = 5$ at fixed $\beta = 1$ and varying 
values of $B$. 
The data are obtained by relying on the class of typical pure quantum 
states introduced in Eq.\ \eqref{Eq::TypicalState}. 
In Appendix 
\ref{Sec::Accu_Typ}, we demonstrate that this DQT approach 
indeed yields accurate results by comparing to exact diagonalization for
smaller system sizes. 
As expected, we find 
that the initial value $\langle M(0)\rangle$ increases 
with increasing $B$ (see insets in Fig.\ \ref{Fig2}).  
Correspondingly, the resulting dynamics $\langle M(t) \rangle$ 
approaches with increasing $B$ the dynamics obtained from the fully polarized 
state 
$\ket{\uparrow}$ (dashed curves).
Moreover, we 
find that for a given $B$, the initial 
value $\langle M(0) \rangle$ is lower for $W = 5$ than for $W = 1$, i.e, at stronger disorder a larger external field 
$B$ is required to polarize $\rho(0)$. 

To better analyze the impact of $B$, 
the main panels in Fig.\ \ref{Fig2} show $\langle M(t) \rangle$ rescaled by their $B$-dependent initial values $\langle M(0)\rangle$. 
Remarkably, we find that for both $W = 1$ and $W = 
5$, the dynamics for $\beta = 1$ and different $B$, as well as for the 
fully polarized state $\ket{\uparrow}$, are all rather
similar to each other. 
The temporal relaxation of the magnetization 
thus appears to be almost independent of whether the initial state is prepared 
close to or far away from equilibrium, see also Refs.\ \cite{Richter_2018, Richter_2019, Richter_2019_2}. 

This apparent independence of the dynamics on the initial state (i.e., the choice of $B$) can be understood especially at strong disorder $W = 5$, for which the system is in a finite-size MBL regime. In this case, there exists a set of (approximate) l-bits and the overlap of these l-bits with $M$ will determine the long-time value of $\langle M(t)\rangle$. This overlap will be almost independent of $B$ such that the long-time value $\langle M(t\to \infty)\rangle$ is a fixed fraction of the initial magnetization. Therefore, when rescaled by $\langle M(0)\rangle$, curves for different $B$ become very similar. On the other hand, such an argument does not immediately apply at weaker disorder $W = 1$. Indeed, the data in Fig.\ \ref{Fig2}~(a) suggests a ``close-to-equilibrium'' regime with the curves for $B = 0.5, 1, 2$ all agreeing perfectly, and a ``far-from-equilibrium'' regime as $\langle M(t)\rangle$ for $B = 4$ and $B \to \infty$ (i.e., $\rho(0) \to \ket{\uparrow}$) appear to be slightly different compared to the lower-$B$ dynamics. This effect is however admittedly quite weak.

\subsection{Driven system}\label{Sec::Results_Drive}

\begin{figure}[tb]
 \centering
 \includegraphics[width=1\columnwidth]{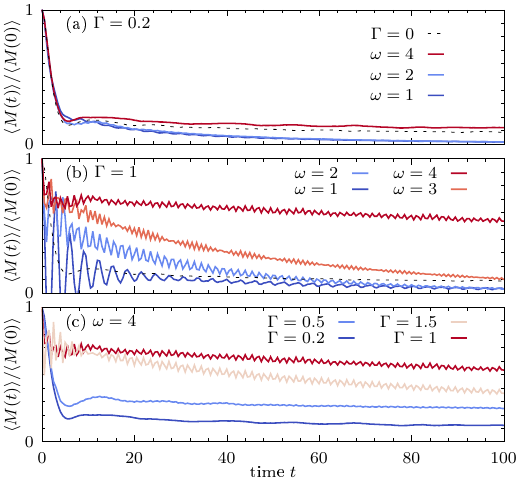}
 \caption{$\langle M(t)\rangle$ starting 
 from the fully polarized initial state 
$\ket{\uparrow}$, time-evolved under 
the driven Hamiltonian $H(t)$ [Eq.\ 
\eqref{Eq::Hdriven}] with weak disorder $W = 1$ and driving strength {\bf 
(a)} $\Gamma = 0.2$; and {\bf (b)} $\Gamma = 1$. Data is shown for different 
driving frequencies $\omega$. The dashed curve indicates the dynamics 
in the undriven case ($\Gamma = 0$). 
{\bf (c)} $\omega = 4$ and varying $\Gamma$. The system size is $L = 18$ in all cases.}
 \label{Fig3}
\end{figure}

We now turn to the dynamics of the driven system $H(t)$ with $\Gamma > 0$ and 
$\omega > 0$. In Fig.\ \ref{Fig3}, we consider a weakly disordered spin chain 
with $W = 1$, for which $\langle M(t) \rangle$ decayed quickly in the undriven case [cf.\ Fig.\ \ref{Fig1}].
Specifically, in Fig.\ \ref{Fig3}~(a), $\langle M(t) \rangle$ is shown for small $\Gamma = 0.2$ (``linear response regime'' \cite{Herbrych_2016}) and different driving frequencies $\omega$. Generally, $\langle M(t) \rangle$ behaves rather similarly to the undriven dynamics (dashed curve for comparison) with a fast decay towards an unpolarized state. While this decay appears to be even facilitated at $\omega = 1,2$, we observe a small amount of induced magnetization for the largest frequency $\omega = 4$ shown here (red curve above dashed curve).   

A considerably stronger effect can be achieved by going beyond the linear-response regime and considering a drive with  amplitude $\Gamma = 1$. In particular, as shown in Fig.\ \ref{Fig3}~(b), we find that at $\omega  = 1,2$ the decay of $\langle M(t) \rangle$ exhibits distinct (damped) oscillations with period $\sim \pi/\omega$. While the response to the drive is strongest at short times, the oscillations die out at later times and the long-time behavior close to equilibrium is almost unchanged compared to the undriven system. This ``stalled'' response near thermal equilibrium has been recently proposed as a more general phenomenon of driven many-body quantum systems \cite{Dabelow_2024}.

With increasing driving frequency $\omega$, the oscillations of $\langle M(t)\rangle$ become less pronounced. More importantly, Fig.\ \ref{Fig3}~(b) unveils that the decay of $\langle M(t)\rangle$ becomes slower with increasing $\omega$, i.e., the drive induces a net magnetization into the system. This effect is particularly striking at $\omega = 4$, where $\langle M(t) \rangle$ at $t = 100$ is still significantly larger compared to the undriven case. While the data in Fig.\ \ref{Fig3} are obtained for system size $L = 18$, we show in Appendix \ref{Sec::App_FS} that the data are essentially converged with respect to $L$.

The dependence on the driving strength is studied in Fig.\ \ref{Fig3}~(c), where $\langle M(t)\rangle$ is shown for various $\Gamma$ at fixed $\omega = 4$. While the induced magnetization increases as expected by increasing the amplitude from $\Gamma = 0.2$ up to $\Gamma = 1$, we find that, somewhat counterintuitively, $\langle M(t) \rangle$ again decays faster when applying an even stronger drive with $\Gamma = 1.5$. Stronger driving thus not necessarily leads to a stronger response, see also \cite{Richter_2019_3} for similar findings. Note that a similar effect is also expected for the dependence of $\langle M(t) \rangle$ on the frequency $\omega$. Specifically, there will be a resonance frequency (here numerically found as $\omega \approx 4$) for which the induced magnetization is largest. Increasing $\omega$ further beyond the resonance frequency (not shown here) will not yield a stronger effect since the system is unable absorb energy from the drive \cite{Abanin_2017, Kuwahara_2016}. 

\begin{figure}[tb]
 \centering
 \includegraphics[width=1\columnwidth]{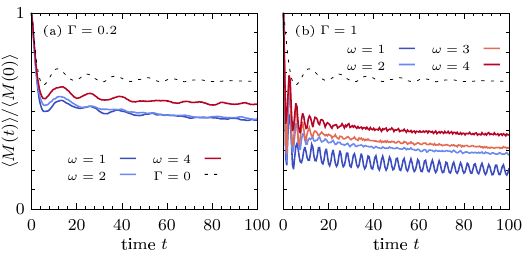}
 \caption{Analogous data as in Fig.\ \ref{Fig3}~(a),(b), but now for stronger disorder 
$W = 5$.}
 \label{Fig3_W5}
\end{figure}

To proceed, in Fig.\ \ref{Fig3_W5}, we study the impact of driving at stronger disorder $W = 5$, where 
the isolated system behaves fairly localized. We again consider weak driving with $\Gamma = 0.2$ and stronger driving with $\Gamma = 1$. We find that the effect of $\Gamma > 0$ is qualitatively different to the weakly disordered case with $W = 1$ considered before. In particular, no excess magnetization is induced compared to the undriven $\Gamma = 0$ case. Rather, $\langle M(t) \rangle$ is reduced compared to the undriven case. While full MBL can suppress drive-induced heating \cite{Khemani_2016}, the data in Fig.\ \ref{Fig3_W5} suggest that the drive facilitates thermalization for all $\omega$ and $\Gamma$ shown here. Furthermore, comparing curves for different $\omega > 0$ in Fig.\ \ref{Fig3_W5}, we find that the curves for increasing $\omega$ approach the undriven $\Gamma = 0$ dynamics from below. This can again be understood from the fact that for sufficiently high $\omega$, the system absorbs less and less energy from the drive.

Building on previous works, where models with single-ion anisotropy were studied \cite{Takayoshi_2014, Takayoshi_2014_2, Herbrych_2016}, we have shown in Figs.\ \ref{Fig3} and \ref{Fig3_W5} that circularly polarized light can be used to alter the dynamics of disordered systems considered in the context of MBL. While the laser-induced magnetization was explained in Refs.\ \cite{Takayoshi_2014, Takayoshi_2014_2} by using a mapping to an effective static model, valid in case that the original time-independent $H$ has a U$(1)$ symmetry and conserves the total magnetization $M$, our numerical simulations show that this phenomenon also occurs for $H$ that do not conserve $M$, see Eq.\ \eqref{Eq::Ham0}. Furthermore, we have demonstrated that for our initial-state protocol, the external driving allows to distinguish between regimes of weak disorder, where excess magnetization is induced to the system, as well as regimes of stronger disorder where the drive leads to a faster relaxation of magnetization. This is a main result. 
While this distinction is pronounced at strong driving with $\Gamma = 1$, we should note that achieving such strong magnetic-field intensities in lasers might be a challenge in experiments \cite{Takayoshi_2014, Takayoshi_2014_2}.  

We note that a similar distinction between weakly and strongly disordered systems can be obtained also in case of the driven isotropic Heisenberg chain with $J^{x,y,z} = 1$, which we demonstrate in Appendix \ref{Sec::App::Heis} using a similar nonequilibrium protocol.

\subsection{(Driven-)dissipative system}\label{Sec::Results_Open}

\begin{figure}[tb]
 \centering
 \includegraphics[width=1\columnwidth]{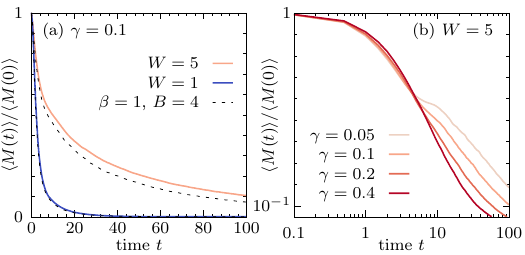}
 \caption{{\bf (a)} Relaxation of magnetization under Lindblad 
 dynamics with $\gamma = 0.1$, and disorder $W = 1$ and $W = 5$. Solid curves indicate data resulting from the pure initial state $\ket{\uparrow}$, while dashed curves result from a mixed initial state with $\beta = 1$ and $B = 4$ obtained using the typicality approach. The system size is $L = 20$ and data are normalized by the initial value $\langle M(0)\rangle$. {\bf (b)} 
Dynamics at fixed $W = 5$ and $L = 18$ for varying system-bath coupling $\gamma$ in a 
logarithmic plot. We have $\Gamma = 0$ (i.e., no driving) in 
all cases.}
 \label{Fig::Open}
\end{figure}

We now also consider the influence of an environment, modelled 
by the Lindblad dynamics in Eqs.\ \eqref{Eq:Lindblad} and \eqref{Eq:Lindblad2} 
with system-bath coupling $\gamma > 0$. 
In Fig.\ \ref{Fig::Open}~(a), the magnetization 
$\langle M(t)\rangle$ is shown at $\gamma = 0.1$, $W = 1$ and $W = 
5$, without external driving ($\Gamma = 0$). 
Compared to the dynamics of the isolated system, $\langle M(t) \rangle$ now decays monotonically towards zero due to the dephasing noise, both for $W = 1$ and $W = 5$, although the relaxation is still slower at stronger disorder.

In Fig.\ \ref{Fig::Open}~(a), we not only show data for the fully polarized initial state $\ket{\psi(0)} = \ket{\uparrow}$, but also for mixed states $\rho(0)$ with $\beta = 1$ and $B = 4$, obtained using stochastic unraveling with the random pure states in Eq.\ \eqref{Eq::TypicalState}. Analogous to the isolated system (Fig.\ \ref{Fig2}), we find that the dynamics resulting from the mixed and pure initial conditions are rather similar to each other. Importantly, we emphasize that the combination of quantum typicality and stochastic unraveling allows us to simulate mixed-state Lindblad dynamics in an open system with $L = 20$, which is clearly beyond the range of exact diagonalization where studies are typically limited to $L < 10$ \cite{O_Sullivan_2020}.

The impact of the environment is further exemplified in Fig.\ \ref{Fig::Open}~(b), where $\langle M(t)\rangle$ is shown at fixed $W = 5$ and 
varying $\gamma$. As can be seen from the logarithmic plot, we can distinguish between two regimes, i.e., $t \lesssim 5$, where curves for different $\gamma$ coincide; and longer times $t \gtrsim 5$, where the decay of $\langle M(t) \rangle$ is more rapid with increasing $\gamma$. The two regimes can be partially understood from the dynamics of the isolated system (Fig.\ \ref{Fig1}), where $\langle M(t)\rangle$ at $W =5$ decays at short times but is approximately constant at longer times. The short-time decay in Fig.\ \ref{Fig::Open} is thus dominated by the internal dynamics of $H$, while the decay at $t\gtrsim 5$ is caused by $\gamma > 0$. 

\begin{figure}[tb]
    \centering
    \includegraphics[width=0.9\columnwidth]{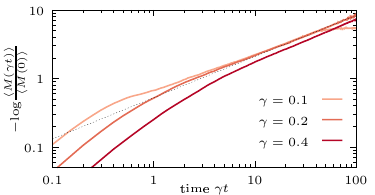}
    \caption{Visualizing the stretched-exponential decay by plotting -$\text{log}\frac{\langle M(t)\rangle}{\langle M(0)\rangle}$ versus rescaled time $\gamma t$. The dashed curve is a fit to the late-time data which can be used to extract the stretching exponent $\alpha$. For $\gamma = 0.1$ and late times, deviations from the stretched-exponential decay are visible.}
    \label{fig:SE}
\end{figure}

The shape of the curves in Fig.\ \ref{Fig::Open}~(b) suggests that the decay of $\langle M(t)\rangle$ is not exponential, but rather described by a stretched-exponential behavior \cite{Fischer_2016, 
Levi_2016, Medvedyeva_2016, Everest_2017, Gopalakrishnan_2017}, 
\begin{equation}\label{Eq::Stretch}
 \langle M(t) \rangle \propto e^{-\lambda t^\alpha}\ ,  
\end{equation}
where $\lambda > 0$ is a constant and $\alpha > 0$ is the stretching exponent.
The scaling \eqref{Eq::Stretch} can be confirmed by plotting $-\text{log}\langle M(t) \rangle \propto \lambda t^\alpha$ in Fig.\ \ref{fig:SE}. Indeed, we find that at sufficiently long times $-\text{log}\langle M(t) \rangle$ grows linearly in the double logarithmic plot and $\alpha$ can be extracted from the slope (dashed curve in Fig.\ \ref{fig:SE}).  
Moreover, we find that the strength of the system-bath coupling $\gamma$ does not have a qualitative effect as the dynamics for different $\gamma$ nicely collapse onto each other when plotted against $t \to \gamma t$. 

\begin{figure}[tb]
 \centering
 \includegraphics[width=1\columnwidth]{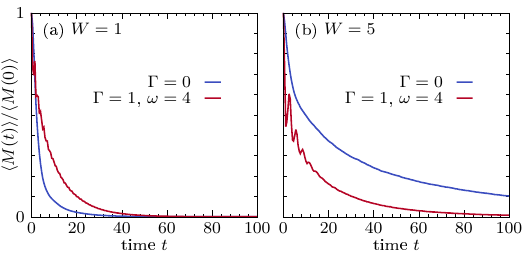}
 \caption{Dynamics in the presence of driving and dissipation 
 for {\bf (a)} weak disorder $W = 1$; and {\bf (b)} stronger disorder $W = 5$. 
 The system-bath coupling is $\gamma = 0.1$.}
 \label{Fig_Open_Drive}
\end{figure}

Eventually, let us study how additional driving impacts the relaxation of $\langle M(t)\rangle$ in the open system (i.e., $\Gamma,\omega,\gamma > 0$). 
In Fig.\ \ref{Fig_Open_Drive}, we fix $\gamma = 0.1$ and consider both, weak disorder $W = 1$ [Fig.\ \ref{Fig_Open_Drive}~(a)] and stronger disorder $W = 5$ [Fig.\ \ref{Fig_Open_Drive}~(b)]. 
Comparing $W = 1$ and $W = 5$, we observe a behavior similar to our discussion earlier in the context of the isolated system. Namely, while the drive slows down relaxation at weak disorder, it appears to further accelerate relaxation in the strongly disordered case. Thus, even in the presence of dephasing, the driving protocol allows to distinguish between a weakly and a strongly disordered regime. Generally, however, the impact of driving, and in particular the enhancement of magnetization at weak disorder, is less striking compared our previous result for the closed system. Observing the laser-induced magnetization experimentally is thus more likely in systems where the spin-spin interactions dominate compared to, e.g., the spin-phonon coupling, such that the dynamics remain approximately unitary on longer time scales.

Figures \ref{Fig::Open} - \ref{Fig_Open_Drive} demonstrate that finite dephasing with $\gamma > 0$ 
facilitates 
the relaxation of $\langle M(t) \rangle$,  
even though the Lindblad jump operators $\sigma_\ell^z$ themselves 
commute with the total magnetization $M$. 
This can be understood from the fact that, since the isolated system $H$ does 
not conserve $M$, and the Lindblad 
dynamics consists of both the unitary and the dissipative part, 
the the only fixed point 
of Eq.\ \eqref{Eq:Lindblad} is the maximally mixed-state $\rho(t\to \infty) 
\propto \mathbb{1}$ such that $\langle M(t\to \infty)\rangle \to 0$. 
Thus, in our setup, any signatures of localization disappear at long times, even in the driven system with $\Gamma,\omega > 0$. Therefore, we have here 
focused on the interplay between disorder, 
driving, and dissipation, on intermediate time scales prior to equilibration 
\cite{Ren_2020}.
Generally, it would be interesting to see if the revival of localization in the steady state due to driving reported in Ref.\ \cite{Lenar_i__2018} can be seen in the transient nonequilibrium dynamics of driven-dissipative quantum systems using other driving protocols or more general non-Markovian system-bath scenarios beyond our setup considered here. 
%


\section{Conclusion}\label{Sec::Conclu}
In summary, we have studied the temporal relaxation of 
magnetization in disordered quantum spin chains focusing on a realistic quench protocol where 
a system is initially polarized due to a strong magnetic field which is subsequently removed to induce a nonequilibrium situation. This protocol was motivated by earlier work in Ref.\ \cite{Ros_2017}, where it was shown that the remanent magnetization of the system at long times can be seen as a proxy for the crossover to a finite-size many-body localized regime. Here, we have particularly studied the impact of driving and dissipation on the resulting dynamics of the system. 

Another motivation for our study was given by Ref.\ \cite{Lenar_i__2018}, which showed 
that fingerprints of MBL can be revived in the steady state of driven systems, even in the presence of a bath that would usually lead to thermalization. 
In contrast, we here focused on the dynamics of disordered driven-dissipative systems on intermediate time scales. To this end, we explored the possibility of inducing magnetization by driving the system with circularly polarized light \cite{Takayoshi_2014, Takayoshi_2014_2, Herbrych_2016}. As a main result, we demonstrated that such a driving protocol indeed allows to distinguish between systems with weak disorder and systems with strong disorder. Specifically, we found that in the strongly disordered case the drive facilitates relaxation and leads to a reduction of the magnetization. In contrast, at weaker disorder and using suitable values of the drive amplitude and frequency, we found that the drive induces a significant amount of excess magnetization such that the system's relaxation becomes slower compared to the undriven case. Let us stress, however, that this slower relaxation should not be interpreted as an onset of localization. Rather, we still 
expect the system to behave thermal, although the thermal value of magnetization is nonzero and set by an appropriate effective Hamiltonian \cite{Luitz_2020}.

Eventually, when we considered additional dephasing noise modeled by a Lindblad master equation, the decay of magnetization towards equilibrium was found to be consistent with stretched-exponential behavior \cite{Fischer_2016, 
Levi_2016, Medvedyeva_2016, Everest_2017, Gopalakrishnan_2017}. While the system's response to driving again revealed differences between weak and strong disorder, these signatures turned out to be less striking than in the unitary case. Moreover, at long times, the system necessarily decayed towards a featureless infinite-temperature state in our setup. 

From a numerical point of view, we used an efficient scheme based on quantum typicality, 
where mixed nonequilibrium initial states can be mimicked by a 
suitably prepared random pure quantum state. Random quantum states have 
been used extensively in previous works to study the 
dynamics of isolated quantum many-body systems (see e.g., \cite{Heitmann_2020, Jin_2021, Elsayed2013, Iitaka2003, Steinigeweg_2016, Richter2019b}). By combining typicality with stochastic unraveling 
of Lindblad master equations, we have here demonstrated
that they also provide a useful numerical tool to study the 
dynamics of open systems prepared in a class of 
experimentally-relevant initial states, cf.\ Refs.\ \cite{Heitmann_2022, Heitmann_2023}. Moreover, while powerful tensor-network algorithms certainly exist to simulate Lindblad dynamics \cite{Zwolak_2004, Verstraete_2004, _nidari__2016, Lenar_i__2020}, the appeal of the combination of typicality and stochastic unraveling lies in its simplicity and its applicability irrespective of details of the system or the jump operators.   

In the future, it would be interesting to study in more detail the differences between internal mechanisms of thermalization and
environment-caused thermalization in potential MBL systems. Distinguishing between these internal and external mechanisms can be challenging since the dynamics of disordered, slowly thermalizing, isolated systems is also of streched-exponential form \cite{Lezama_2019, Crowley_2022, Long2022}.
Another natural avenue is to further explore potential applications of quantum  typicality in the context of open quantum systems, e.g., in order to study the 
dynamics of non-Hermitian Hamiltonians \cite{Mahoney2024}. Finally, it would be 
interesting to see if the here reported response of disordered spin chains can 
be experimentally observed either in solid-state settings or noisy quantum 
simulators.  

\subsection*{Acknowledgements}
We thank Vedika Khemani, Yaodong Li, and Alan Morningstar for helpful comments and discussions. Moreover, we thank Jochen Gemmer, Jacek Herbrych, and Robin Steinigeweg for stimulating discussions and previous collaborations on related topics.
J.\,R.\ acknowledges funding from the European Union's Horizon Europe research 
and innovation programme, Marie Sk\l odowska-Curie grant no.\ 101060162. J.\,R.\ is also supported by the Packard Foundation through a
Packard Fellowship in Science and Engineering (V.\, Khemani's grant). 


\appendix

\section{Decay of magnetization in the disordered isotropic Heisenberg chain}\label{Sec::App::Heis}

While we have focused on anisotropic couplings with $J^y = 0$ in the main text, we here present 
additional results for the isotropic Heisenberg chain with $J^{x,y,z} = 1$. 
In contrast to our analysis in the main text, the undriven $H$ therefore now conserves the total magnetization $M$. 
Due to a strong external magnetic field, the system is again prepared in the fully polarized initial state $\rho(0) = \ket{\psi(0)}\bra{\psi(0)}$ with $\ket{\psi(0)} = \ket{\uparrow \uparrow \cdots \uparrow}$. At time $t = 0$, the external field is removed and the system is driven by circularly polarized light of amplitude $\Gamma$ and frequency $\omega$, cf.\ Eq.\ \eqref{Eq::DriveDrive}. Due to the drive, $M$ is not conserved under the time-dependent $H(t)$ such that $\langle M(t) \rangle$ shows nontrivial dynamics. 
\begin{figure}[b]
 \centering
 \includegraphics[width=0.95\columnwidth]{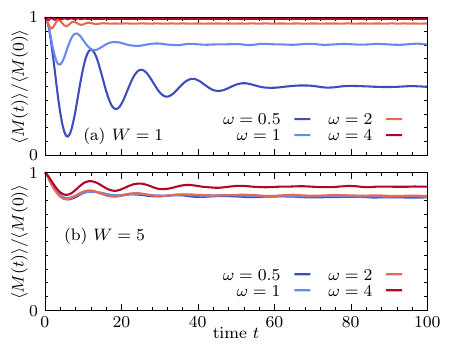}
 \caption{Decay of magnetization in the isotropic Heisenberg chain with $J^x = J^y = J^z = 1$. The system is driven in the linear-response regime with driving amplitude $\Gamma = 0.2$ and data are shown for various driving frequencies $\omega$. The disorder strength is {\bf (a)} $W = 1$, and {\bf (b)} $W = 5$. We have $L = 18$ in all cases.}
 \label{FigS_XXZ}
\end{figure}

In Fig.\ \ref{FigS_XXZ}, $\langle M(t) \rangle$ is shown for both weak disorder $W = 1$ [Fig.\ \ref{FigS_XXZ}~(a)] and strong disorder $W = 5$ [Fig.\ \ref{FigS_XXZ}~(b)]. We focus on the linear-response regime with $\Gamma = 0.2$ and study the relaxation of magnetization for various driving frequencies $\omega$. Interestingly, there is a clear difference in the system's response when comparing the two disorder strengths. On one hand, for $W = 1$ and $\omega = 0.5$, we find that $\langle M(t)\rangle$ exhibits distinct oscillations and decays to a notably reduced long-time value. With increasing $\omega$, the oscillations become less pronounced and the long-time value of $\langle M(t)\rangle$ increases. In particular, for high driving frequency $\omega = 4$, $\langle M(t)\rangle$ is again approximately conserved with $\langle M(t) \rangle/\langle M(0)\rangle \approx 1$ for all times shown here.

On the other hand, for stronger disorder $W = 5$, the dependence of $\langle M(t) \rangle$ on the choice of $\omega$ is significantly weaker. In fact, we find that the dynamics remain essentially unchanged when varying $\omega$ from $\omega = 0.5$ to $\omega = 2$. The data in Fig.\ \ref{FigS_XXZ} thus exemplifies that driving a disordered spin chain by circularly polarized light, together with the resulting response of the nonequilibrium magnetization, allows to distinguish between strongly disordered (which show a weak response) and weakly disordered systems (which show a strong response). While we have demonstrated this finding in the main text for systems where the undriven $H$ does not conserve $M$, Fig.\ \ref{FigS_XXZ} shows that a similar protocol also applies in the case of isotropic couplings where $H$ has a U$(1)$ symmetry.

\section{Finite-size scaling of dynamics at weak disorder}\label{Sec::App_FS}

\begin{figure}[t]
 \centering
 \includegraphics[width=0.95\columnwidth]{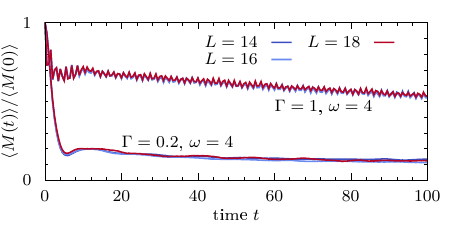}
  \caption{Finite-size scaling analysis for setup analogous to Fig.\  \ref{Fig3}, using a fixed driving frequency $\omega = 4$. Data is shown for different 
 system sizes $L = 14,16,18$ and driving strength $\Gamma = 0.2$ and  $\Gamma = 1$.}
 \label{FigS4}
\end{figure}

In Fig.\ \ref{FigS4}, we show the relaxation 
of magnetization in driven spin chains with $\Gamma = 0.2$ and $\Gamma = 1$ at 
weak disorder $W = 1$ (i.e., analogous to Fig.\ \ref{Fig3} in the main text). The driving frequency is $\omega 
= 4$. 
Plotting data for system sizes $L = 14,16,18$, we find that curves of 
$\langle M(t)\rangle/\langle M(0)\rangle$ with different $L$ almost 
perfectly coincide with each other, i.e., finite-size effects are negligible on the 
time scales shown here.
\begin{figure}[b]
 \centering
 \includegraphics[width=0.95\columnwidth]{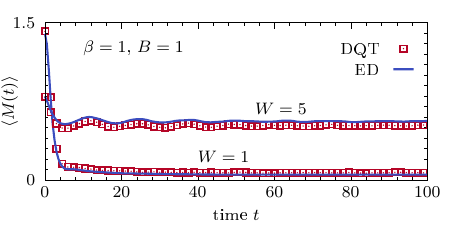}
 \caption{Comparison of dynamical quantum typicality (DQT) and exact 
diagonalization (ED). The mixed initial 
state $\rho(0)$ is prepared with $\beta = 1$ and $B = 1$ 
and we study the unitary time evolution governed by the isolated 
model $H$ with {\bf (a)} $W = 1$; and {\bf (b)} $W = 5$. The system size 
is $L = 10$.}
 \label{FigS1}
\end{figure}

\section{Accuracy of dynamical quantum typicality}\label{Sec::Accu_Typ}

Let us exemplify that the random pure states introduced in Eq.\ 
\eqref{Eq::TypicalState} indeed yield accurate results for the relaxation of 
$\langle M(t) \rangle$ in the case of mixed initial states $\rho(0)$ with 
finite $\beta$ and $B$. To this end, Fig.\ \ref{FigS1} shows $\langle 
M(t)\rangle$ in small chains with $L = 10$ for weak disorder $W = 1$ and strong disorder $W = 5$. In both cases, we
compare data obtained by dynamical quantum typicality to results obtained from 
exact diagonalization. Note that for the DQT data, averaging over random states is performed simultaneously with averaging over disorder realizations. 

Generally, Fig.\ \ref{FigS1} unveils a convincing agreement between DQT and ED even for the small system size used. While this agreement is of slightly lower quality for $W = 5$, we note that this is caused by the fact that sample-to-sample variations are larger at stronger disorder and that ED and DQT data are here obtained for different sets of random disorder realizations.


\begin{thebibliography}{77}%
\makeatletter
\providecommand \@ifxundefined [1]{%
 \@ifx{#1\undefined}
}%
\providecommand \@ifnum [1]{%
 \ifnum #1\expandafter \@firstoftwo
 \else \expandafter \@secondoftwo
 \fi
}%
\providecommand \@ifx [1]{%
 \ifx #1\expandafter \@firstoftwo
 \else \expandafter \@secondoftwo
 \fi
}%
\providecommand \natexlab [1]{#1}%
\providecommand \enquote  [1]{``#1''}%
\providecommand \bibnamefont  [1]{#1}%
\providecommand \bibfnamefont [1]{#1}%
\providecommand \citenamefont [1]{#1}%
\providecommand \href@noop [0]{\@secondoftwo}%
\providecommand \href [0]{\begingroup \@sanitize@url \@href}%
\providecommand \@href[1]{\@@startlink{#1}\@@href}%
\providecommand \@@href[1]{\endgroup#1\@@endlink}%
\providecommand \@sanitize@url [0]{\catcode `\\12\catcode `\$12\catcode
  `\&12\catcode `\#12\catcode `\^12\catcode `\_12\catcode `\%12\relax}%
\providecommand \@@startlink[1]{}%
\providecommand \@@endlink[0]{}%
\providecommand \url  [0]{\begingroup\@sanitize@url \@url }%
\providecommand \@url [1]{\endgroup\@href {#1}{\urlprefix }}%
\providecommand \urlprefix  [0]{URL }%
\providecommand \Eprint [0]{\href }%
\providecommand \doibase [0]{https://doi.org/}%
\providecommand \selectlanguage [0]{\@gobble}%
\providecommand \bibinfo  [0]{\@secondoftwo}%
\providecommand \bibfield  [0]{\@secondoftwo}%
\providecommand \translation [1]{[#1]}%
\providecommand \BibitemOpen [0]{}%
\providecommand \bibitemStop [0]{}%
\providecommand \bibitemNoStop [0]{.\EOS\space}%
\providecommand \EOS [0]{\spacefactor3000\relax}%
\providecommand \BibitemShut  [1]{\csname bibitem#1\endcsname}%
\let\auto@bib@innerbib\@empty
\bibitem [{\citenamefont {Eisert}\ \emph {et~al.}(2015)\citenamefont {Eisert},
  \citenamefont {Friesdorf},\ and\ \citenamefont {Gogolin}}]{Eisert_2015}%
  \BibitemOpen
  \bibfield  {author} {\bibinfo {author} {\bibfnamefont {J.}~\bibnamefont
  {Eisert}}, \bibinfo {author} {\bibfnamefont {M.}~\bibnamefont {Friesdorf}},\
  and\ \bibinfo {author} {\bibfnamefont {C.}~\bibnamefont {Gogolin}},\ }\href
  {https://doi.org/https://doi.org/10.1038/nphys3215} {\bibfield  {journal}
  {\bibinfo  {journal} {Nature Physics}\ }\textbf {\bibinfo {volume} {11}},\
  \bibinfo {pages} {124} (\bibinfo {year} {2015})}\BibitemShut {NoStop}%
\bibitem [{\citenamefont {Nandkishore}\ and\ \citenamefont
  {Huse}(2015)}]{Nandkishore_2015}%
  \BibitemOpen
  \bibfield  {author} {\bibinfo {author} {\bibfnamefont {R.}~\bibnamefont
  {Nandkishore}}\ and\ \bibinfo {author} {\bibfnamefont {D.~A.}\ \bibnamefont
  {Huse}},\ }\href
  {https://doi.org/https://doi.org/10.1146/annurev-conmatphys-031214-014726}
  {\bibfield  {journal} {\bibinfo  {journal} {Annual Review of Condensed Matter
  Physics}\ }\textbf {\bibinfo {volume} {6}},\ \bibinfo {pages} {15} (\bibinfo
  {year} {2015})}\BibitemShut {NoStop}%
\bibitem [{\citenamefont {Abanin}\ \emph {et~al.}(2019)\citenamefont {Abanin},
  \citenamefont {Altman}, \citenamefont {Bloch},\ and\ \citenamefont
  {Serbyn}}]{Abanin_2019}%
  \BibitemOpen
  \bibfield  {author} {\bibinfo {author} {\bibfnamefont {D.~A.}\ \bibnamefont
  {Abanin}}, \bibinfo {author} {\bibfnamefont {E.}~\bibnamefont {Altman}},
  \bibinfo {author} {\bibfnamefont {I.}~\bibnamefont {Bloch}},\ and\ \bibinfo
  {author} {\bibfnamefont {M.}~\bibnamefont {Serbyn}},\ }\href
  {https://doi.org/https://doi.org/10.1103/RevModPhys.91.021001} {\bibfield
  {journal} {\bibinfo  {journal} {Reviews of Modern Physics}\ }\textbf
  {\bibinfo {volume} {91}},\ \bibinfo {pages} {021001} (\bibinfo {year}
  {2019})}\BibitemShut {NoStop}%
\bibitem [{\citenamefont {Pal}\ and\ \citenamefont {Huse}(2010)}]{Pal_2010}%
  \BibitemOpen
  \bibfield  {author} {\bibinfo {author} {\bibfnamefont {A.}~\bibnamefont
  {Pal}}\ and\ \bibinfo {author} {\bibfnamefont {D.~A.}\ \bibnamefont {Huse}},\
  }\href {https://doi.org/https://doi.org/10.1103/PhysRevB.82.174411}
  {\bibfield  {journal} {\bibinfo  {journal} {Physical Review B}\ }\textbf
  {\bibinfo {volume} {82}},\ \bibinfo {pages} {174411} (\bibinfo {year}
  {2010})}\BibitemShut {NoStop}%
\bibitem [{\citenamefont {Kj{\"a}ll}\ \emph {et~al.}(2014)\citenamefont
  {Kj{\"a}ll}, \citenamefont {Bardarson},\ and\ \citenamefont
  {Pollmann}}]{Kj_ll_2014}%
  \BibitemOpen
  \bibfield  {author} {\bibinfo {author} {\bibfnamefont {J.~A.}\ \bibnamefont
  {Kj{\"a}ll}}, \bibinfo {author} {\bibfnamefont {J.~H.}\ \bibnamefont
  {Bardarson}},\ and\ \bibinfo {author} {\bibfnamefont {F.}~\bibnamefont
  {Pollmann}},\ }\href
  {https://doi.org/https://doi.org/10.1103/PhysRevLett.113.107204} {\bibfield
  {journal} {\bibinfo  {journal} {Physical Review Letters}\ }\textbf {\bibinfo
  {volume} {113}},\ \bibinfo {pages} {107204} (\bibinfo {year}
  {2014})}\BibitemShut {NoStop}%
\bibitem [{\citenamefont {Luitz}\ \emph {et~al.}(2015)\citenamefont {Luitz},
  \citenamefont {Laflorencie},\ and\ \citenamefont {Alet}}]{Luitz_2015}%
  \BibitemOpen
  \bibfield  {author} {\bibinfo {author} {\bibfnamefont {D.~J.}\ \bibnamefont
  {Luitz}}, \bibinfo {author} {\bibfnamefont {N.}~\bibnamefont {Laflorencie}},\
  and\ \bibinfo {author} {\bibfnamefont {F.}~\bibnamefont {Alet}},\ }\href
  {https://doi.org/https://doi.org/10.1103/PhysRevB.91.081103} {\bibfield
  {journal} {\bibinfo  {journal} {Physical Review B}\ }\textbf {\bibinfo
  {volume} {91}},\ \bibinfo {pages} {081103} (\bibinfo {year}
  {2015})}\BibitemShut {NoStop}%
\bibitem [{\citenamefont {Imbrie}(2016)}]{Imbrie_2016}%
  \BibitemOpen
  \bibfield  {author} {\bibinfo {author} {\bibfnamefont {J.~Z.}\ \bibnamefont
  {Imbrie}},\ }\href
  {https://doi.org/https://doi.org/10.1103/PhysRevLett.117.027201} {\bibfield
  {journal} {\bibinfo  {journal} {Physical Review Letters}\ }\textbf {\bibinfo
  {volume} {117}},\ \bibinfo {pages} {027201} (\bibinfo {year}
  {2016})}\BibitemShut {NoStop}%
\bibitem [{\citenamefont {{\v{S}}untajs}\ \emph {et~al.}(2020)\citenamefont
  {{\v{S}}untajs}, \citenamefont {Bon{\v{c}}a}, \citenamefont {Prosen},\ and\
  \citenamefont {Vidmar}}]{_untajs_2020}%
  \BibitemOpen
  \bibfield  {author} {\bibinfo {author} {\bibfnamefont {J.}~\bibnamefont
  {{\v{S}}untajs}}, \bibinfo {author} {\bibfnamefont {J.}~\bibnamefont
  {Bon{\v{c}}a}}, \bibinfo {author} {\bibfnamefont {T.}~\bibnamefont
  {Prosen}},\ and\ \bibinfo {author} {\bibfnamefont {L.}~\bibnamefont
  {Vidmar}},\ }\href
  {https://doi.org/https://doi.org/10.1103/PhysRevE.102.062144} {\bibfield
  {journal} {\bibinfo  {journal} {Physical Review E}\ }\textbf {\bibinfo
  {volume} {102}},\ \bibinfo {pages} {062144} (\bibinfo {year}
  {2020})}\BibitemShut {NoStop}%
\bibitem [{\citenamefont {Sels}\ and\ \citenamefont
  {Polkovnikov}(2021)}]{Sels_2021}%
  \BibitemOpen
  \bibfield  {author} {\bibinfo {author} {\bibfnamefont {D.}~\bibnamefont
  {Sels}}\ and\ \bibinfo {author} {\bibfnamefont {A.}~\bibnamefont
  {Polkovnikov}},\ }\href
  {https://doi.org/https://doi.org/10.1103/PhysRevE.104.054105} {\bibfield
  {journal} {\bibinfo  {journal} {Physical Review E}\ }\textbf {\bibinfo
  {volume} {104}},\ \bibinfo {pages} {054105} (\bibinfo {year}
  {2021})}\BibitemShut {NoStop}%
\bibitem [{\citenamefont {Morningstar}\ \emph {et~al.}(2022)\citenamefont
  {Morningstar}, \citenamefont {Colmenarez}, \citenamefont {Khemani},
  \citenamefont {Luitz},\ and\ \citenamefont {Huse}}]{Morningstar_2022}%
  \BibitemOpen
  \bibfield  {author} {\bibinfo {author} {\bibfnamefont {A.}~\bibnamefont
  {Morningstar}}, \bibinfo {author} {\bibfnamefont {L.}~\bibnamefont
  {Colmenarez}}, \bibinfo {author} {\bibfnamefont {V.}~\bibnamefont {Khemani}},
  \bibinfo {author} {\bibfnamefont {D.~J.}\ \bibnamefont {Luitz}},\ and\
  \bibinfo {author} {\bibfnamefont {D.~A.}\ \bibnamefont {Huse}},\ }\href
  {https://doi.org/https://doi.org/10.1103/PhysRevB.105.174205} {\bibfield
  {journal} {\bibinfo  {journal} {Physical Review B}\ }\textbf {\bibinfo
  {volume} {105}},\ \bibinfo {pages} {174205} (\bibinfo {year}
  {2022})}\BibitemShut {NoStop}%
\bibitem [{\citenamefont {Abanin}\ \emph {et~al.}(2021)\citenamefont {Abanin},
  \citenamefont {Bardarson}, \citenamefont {Tomasi}, \citenamefont
  {Gopalakrishnan}, \citenamefont {Khemani}, \citenamefont {Parameswaran},
  \citenamefont {Pollmann}, \citenamefont {Potter}, \citenamefont {Serbyn},\
  and\ \citenamefont {Vasseur}}]{Abanin_2021}%
  \BibitemOpen
  \bibfield  {author} {\bibinfo {author} {\bibfnamefont {D.}~\bibnamefont
  {Abanin}}, \bibinfo {author} {\bibfnamefont {J.}~\bibnamefont {Bardarson}},
  \bibinfo {author} {\bibfnamefont {G.~D.}\ \bibnamefont {Tomasi}}, \bibinfo
  {author} {\bibfnamefont {S.}~\bibnamefont {Gopalakrishnan}}, \bibinfo
  {author} {\bibfnamefont {V.}~\bibnamefont {Khemani}}, \bibinfo {author}
  {\bibfnamefont {S.}~\bibnamefont {Parameswaran}}, \bibinfo {author}
  {\bibfnamefont {F.}~\bibnamefont {Pollmann}}, \bibinfo {author}
  {\bibfnamefont {A.}~\bibnamefont {Potter}}, \bibinfo {author} {\bibfnamefont
  {M.}~\bibnamefont {Serbyn}},\ and\ \bibinfo {author} {\bibfnamefont
  {R.}~\bibnamefont {Vasseur}},\ }\href
  {https://doi.org/https://doi.org/10.1016/j.aop.2021.168415} {\bibfield
  {journal} {\bibinfo  {journal} {Annals of Physics}\ }\textbf {\bibinfo
  {volume} {427}},\ \bibinfo {pages} {168415} (\bibinfo {year}
  {2021})}\BibitemShut {NoStop}%
\bibitem [{\citenamefont {Doggen}\ \emph {et~al.}(2018)\citenamefont {Doggen},
  \citenamefont {Schindler}, \citenamefont {Tikhonov}, \citenamefont {Mirlin},
  \citenamefont {Neupert}, \citenamefont {Polyakov},\ and\ \citenamefont
  {Gornyi}}]{Doggen_2018}%
  \BibitemOpen
  \bibfield  {author} {\bibinfo {author} {\bibfnamefont {E.~V.~H.}\
  \bibnamefont {Doggen}}, \bibinfo {author} {\bibfnamefont {F.}~\bibnamefont
  {Schindler}}, \bibinfo {author} {\bibfnamefont {K.~S.}\ \bibnamefont
  {Tikhonov}}, \bibinfo {author} {\bibfnamefont {A.~D.}\ \bibnamefont
  {Mirlin}}, \bibinfo {author} {\bibfnamefont {T.}~\bibnamefont {Neupert}},
  \bibinfo {author} {\bibfnamefont {D.~G.}\ \bibnamefont {Polyakov}},\ and\
  \bibinfo {author} {\bibfnamefont {I.~V.}\ \bibnamefont {Gornyi}},\ }\href
  {https://doi.org/https://doi.org/10.1103/PhysRevB.98.174202} {\bibfield
  {journal} {\bibinfo  {journal} {Physical Review B}\ }\textbf {\bibinfo
  {volume} {98}},\ \bibinfo {pages} {174202} (\bibinfo {year}
  {2018})}\BibitemShut {NoStop}%
\bibitem [{\citenamefont {Panda}\ \emph {et~al.}(2020)\citenamefont {Panda},
  \citenamefont {Scardicchio}, \citenamefont {Schulz}, \citenamefont {Taylor},\
  and\ \citenamefont {{\v{Z}}nidari{\v{c}}}}]{Panda_2020}%
  \BibitemOpen
  \bibfield  {author} {\bibinfo {author} {\bibfnamefont {R.~K.}\ \bibnamefont
  {Panda}}, \bibinfo {author} {\bibfnamefont {A.}~\bibnamefont {Scardicchio}},
  \bibinfo {author} {\bibfnamefont {M.}~\bibnamefont {Schulz}}, \bibinfo
  {author} {\bibfnamefont {S.~R.}\ \bibnamefont {Taylor}},\ and\ \bibinfo
  {author} {\bibfnamefont {M.}~\bibnamefont {{\v{Z}}nidari{\v{c}}}},\ }\href
  {https://doi.org/https://doi.org/10.1209/0295-5075/128/67003} {\bibfield
  {journal} {\bibinfo  {journal} {{EPL} (Europhysics Letters)}\ }\textbf
  {\bibinfo {volume} {128}},\ \bibinfo {pages} {67003} (\bibinfo {year}
  {2020})}\BibitemShut {NoStop}%
\bibitem [{\citenamefont {Richter}\ and\ \citenamefont
  {Pal}(2022)}]{Richter_2022}%
  \BibitemOpen
  \bibfield  {author} {\bibinfo {author} {\bibfnamefont {J.}~\bibnamefont
  {Richter}}\ and\ \bibinfo {author} {\bibfnamefont {A.}~\bibnamefont {Pal}},\
  }\href {https://doi.org/https://doi.org/10.1103/PhysRevB.105.L220405}
  {\bibfield  {journal} {\bibinfo  {journal} {Physical Review B}\ }\textbf
  {\bibinfo {volume} {105}},\ \bibinfo {pages} {l220405} (\bibinfo {year}
  {2022})}\BibitemShut {NoStop}%
\bibitem [{\citenamefont {yoon Choi}\ \emph {et~al.}(2016)\citenamefont {yoon
  Choi}, \citenamefont {Hild}, \citenamefont {Zeiher}, \citenamefont
  {Schau{\ss}}, \citenamefont {Rubio-Abadal}, \citenamefont {Yefsah},
  \citenamefont {Khemani}, \citenamefont {Huse}, \citenamefont {Bloch},\ and\
  \citenamefont {Gross}}]{Choi_2016}%
  \BibitemOpen
  \bibfield  {author} {\bibinfo {author} {\bibfnamefont {J.}~\bibnamefont {yoon
  Choi}}, \bibinfo {author} {\bibfnamefont {S.}~\bibnamefont {Hild}}, \bibinfo
  {author} {\bibfnamefont {J.}~\bibnamefont {Zeiher}}, \bibinfo {author}
  {\bibfnamefont {P.}~\bibnamefont {Schau{\ss}}}, \bibinfo {author}
  {\bibfnamefont {A.}~\bibnamefont {Rubio-Abadal}}, \bibinfo {author}
  {\bibfnamefont {T.}~\bibnamefont {Yefsah}}, \bibinfo {author} {\bibfnamefont
  {V.}~\bibnamefont {Khemani}}, \bibinfo {author} {\bibfnamefont {D.~A.}\
  \bibnamefont {Huse}}, \bibinfo {author} {\bibfnamefont {I.}~\bibnamefont
  {Bloch}},\ and\ \bibinfo {author} {\bibfnamefont {C.}~\bibnamefont {Gross}},\
  }\href {https://doi.org/https://doi.org/10.1126/science.aaf8834} {\bibfield
  {journal} {\bibinfo  {journal} {Science}\ }\textbf {\bibinfo {volume}
  {352}},\ \bibinfo {pages} {1547} (\bibinfo {year} {2016})}\BibitemShut
  {NoStop}%
\bibitem [{\citenamefont {Schreiber}\ \emph {et~al.}(2015)\citenamefont
  {Schreiber}, \citenamefont {Hodgman}, \citenamefont {Bordia}, \citenamefont
  {L{\"u}schen}, \citenamefont {Fischer}, \citenamefont {Vosk}, \citenamefont
  {Altman}, \citenamefont {Schneider},\ and\ \citenamefont
  {Bloch}}]{Schreiber_2015}%
  \BibitemOpen
  \bibfield  {author} {\bibinfo {author} {\bibfnamefont {M.}~\bibnamefont
  {Schreiber}}, \bibinfo {author} {\bibfnamefont {S.~S.}\ \bibnamefont
  {Hodgman}}, \bibinfo {author} {\bibfnamefont {P.}~\bibnamefont {Bordia}},
  \bibinfo {author} {\bibfnamefont {H.~P.}\ \bibnamefont {L{\"u}schen}},
  \bibinfo {author} {\bibfnamefont {M.~H.}\ \bibnamefont {Fischer}}, \bibinfo
  {author} {\bibfnamefont {R.}~\bibnamefont {Vosk}}, \bibinfo {author}
  {\bibfnamefont {E.}~\bibnamefont {Altman}}, \bibinfo {author} {\bibfnamefont
  {U.}~\bibnamefont {Schneider}},\ and\ \bibinfo {author} {\bibfnamefont
  {I.}~\bibnamefont {Bloch}},\ }\href
  {https://doi.org/https://doi.org/10.1126/science.aaa7432} {\bibfield
  {journal} {\bibinfo  {journal} {Science}\ }\textbf {\bibinfo {volume}
  {349}},\ \bibinfo {pages} {842} (\bibinfo {year} {2015})}\BibitemShut
  {NoStop}%
\bibitem [{\citenamefont {Smith}\ \emph {et~al.}(2016)\citenamefont {Smith},
  \citenamefont {Lee}, \citenamefont {Richerme}, \citenamefont {Neyenhuis},
  \citenamefont {Hess}, \citenamefont {Hauke}, \citenamefont {Heyl},
  \citenamefont {Huse},\ and\ \citenamefont {Monroe}}]{Smith_2016}%
  \BibitemOpen
  \bibfield  {author} {\bibinfo {author} {\bibfnamefont {J.}~\bibnamefont
  {Smith}}, \bibinfo {author} {\bibfnamefont {A.}~\bibnamefont {Lee}}, \bibinfo
  {author} {\bibfnamefont {P.}~\bibnamefont {Richerme}}, \bibinfo {author}
  {\bibfnamefont {B.}~\bibnamefont {Neyenhuis}}, \bibinfo {author}
  {\bibfnamefont {P.~W.}\ \bibnamefont {Hess}}, \bibinfo {author}
  {\bibfnamefont {P.}~\bibnamefont {Hauke}}, \bibinfo {author} {\bibfnamefont
  {M.}~\bibnamefont {Heyl}}, \bibinfo {author} {\bibfnamefont {D.~A.}\
  \bibnamefont {Huse}},\ and\ \bibinfo {author} {\bibfnamefont
  {C.}~\bibnamefont {Monroe}},\ }\href
  {https://doi.org/https://doi.org/10.1038/nphys3783} {\bibfield  {journal}
  {\bibinfo  {journal} {Nature Physics}\ }\textbf {\bibinfo {volume} {12}},\
  \bibinfo {pages} {907} (\bibinfo {year} {2016})}\BibitemShut {NoStop}%
\bibitem [{\citenamefont {Ovadia}\ \emph {et~al.}(2015)\citenamefont {Ovadia},
  \citenamefont {Kalok}, \citenamefont {Tamir}, \citenamefont {Mitra},
  \citenamefont {Sac{\'{e}}p{\'{e}}},\ and\ \citenamefont
  {Shahar}}]{Ovadia_2015}%
  \BibitemOpen
  \bibfield  {author} {\bibinfo {author} {\bibfnamefont {M.}~\bibnamefont
  {Ovadia}}, \bibinfo {author} {\bibfnamefont {D.}~\bibnamefont {Kalok}},
  \bibinfo {author} {\bibfnamefont {I.}~\bibnamefont {Tamir}}, \bibinfo
  {author} {\bibfnamefont {S.}~\bibnamefont {Mitra}}, \bibinfo {author}
  {\bibfnamefont {B.}~\bibnamefont {Sac{\'{e}}p{\'{e}}}},\ and\ \bibinfo
  {author} {\bibfnamefont {D.}~\bibnamefont {Shahar}},\ }\href
  {https://doi.org/https://doi.org/10.1038/srep13503} {\bibfield  {journal}
  {\bibinfo  {journal} {Scientific Reports}\ }\textbf {\bibinfo {volume} {5}},\
  \bibinfo {pages} {13503} (\bibinfo {year} {2015})}\BibitemShut {NoStop}%
\bibitem [{\citenamefont {Nietner}\ \emph {et~al.}(2022)\citenamefont
  {Nietner}, \citenamefont {Kshetrimayum}, \citenamefont {Eisert},\ and\
  \citenamefont {Lake}}]{Nietner2022}%
  \BibitemOpen
  \bibfield  {author} {\bibinfo {author} {\bibfnamefont {A.}~\bibnamefont
  {Nietner}}, \bibinfo {author} {\bibfnamefont {A.}~\bibnamefont
  {Kshetrimayum}}, \bibinfo {author} {\bibfnamefont {J.}~\bibnamefont
  {Eisert}},\ and\ \bibinfo {author} {\bibfnamefont {B.}~\bibnamefont {Lake}}\
  }\href {https://doi.org/https://doi.org/10.48550/arXiv.2207.10696}
  {https://doi.org/10.48550/arXiv.2207.10696} (\bibinfo {year}
  {2022})\BibitemShut {NoStop}%
\bibitem [{\citenamefont {Serbyn}\ \emph {et~al.}(2013)\citenamefont {Serbyn},
  \citenamefont {Papi{\'{c}}},\ and\ \citenamefont {Abanin}}]{Serbyn_2013}%
  \BibitemOpen
  \bibfield  {author} {\bibinfo {author} {\bibfnamefont {M.}~\bibnamefont
  {Serbyn}}, \bibinfo {author} {\bibfnamefont {Z.}~\bibnamefont
  {Papi{\'{c}}}},\ and\ \bibinfo {author} {\bibfnamefont {D.~A.}\ \bibnamefont
  {Abanin}},\ }\href
  {https://doi.org/https://doi.org/10.1103/PhysRevLett.111.127201} {\bibfield
  {journal} {\bibinfo  {journal} {Physical Review Letters}\ }\textbf {\bibinfo
  {volume} {111}},\ \bibinfo {pages} {127201} (\bibinfo {year}
  {2013})}\BibitemShut {NoStop}%
\bibitem [{\citenamefont {Huse}\ \emph {et~al.}(2014)\citenamefont {Huse},
  \citenamefont {Nandkishore},\ and\ \citenamefont {Oganesyan}}]{Huse_2014}%
  \BibitemOpen
  \bibfield  {author} {\bibinfo {author} {\bibfnamefont {D.~A.}\ \bibnamefont
  {Huse}}, \bibinfo {author} {\bibfnamefont {R.}~\bibnamefont {Nandkishore}},\
  and\ \bibinfo {author} {\bibfnamefont {V.}~\bibnamefont {Oganesyan}},\ }\href
  {https://doi.org/https://doi.org/10.1103/PhysRevB.90.174202} {\bibfield
  {journal} {\bibinfo  {journal} {Physical Review B}\ }\textbf {\bibinfo
  {volume} {90}},\ \bibinfo {pages} {174202} (\bibinfo {year}
  {2014})}\BibitemShut {NoStop}%
\bibitem [{\citenamefont {Chandran}\ \emph {et~al.}(2015)\citenamefont
  {Chandran}, \citenamefont {Kim}, \citenamefont {Vidal},\ and\ \citenamefont
  {Abanin}}]{Chandran_2015}%
  \BibitemOpen
  \bibfield  {author} {\bibinfo {author} {\bibfnamefont {A.}~\bibnamefont
  {Chandran}}, \bibinfo {author} {\bibfnamefont {I.~H.}\ \bibnamefont {Kim}},
  \bibinfo {author} {\bibfnamefont {G.}~\bibnamefont {Vidal}},\ and\ \bibinfo
  {author} {\bibfnamefont {D.~A.}\ \bibnamefont {Abanin}},\ }\href
  {https://doi.org/https://doi.org/10.1103/PhysRevB.91.085425} {\bibfield
  {journal} {\bibinfo  {journal} {Physical Review B}\ }\textbf {\bibinfo
  {volume} {91}},\ \bibinfo {pages} {085425} (\bibinfo {year}
  {2015})}\BibitemShut {NoStop}%
\bibitem [{\citenamefont {Ros}\ \emph {et~al.}(2015)\citenamefont {Ros},
  \citenamefont {M\"uller},\ and\ \citenamefont {Scardicchio}}]{Ros_2015}%
  \BibitemOpen
  \bibfield  {author} {\bibinfo {author} {\bibfnamefont {V.}~\bibnamefont
  {Ros}}, \bibinfo {author} {\bibfnamefont {M.}~\bibnamefont {M\"uller}},\ and\
  \bibinfo {author} {\bibfnamefont {A.}~\bibnamefont {Scardicchio}},\ }\href
  {https://doi.org/https://doi.org/10.1016/j.nuclphysb.2014.12.014} {\bibfield
  {journal} {\bibinfo  {journal} {Nuclear Physics B}\ }\textbf {\bibinfo
  {volume} {891}},\ \bibinfo {pages} {420} (\bibinfo {year}
  {2015})}\BibitemShut {NoStop}%
\bibitem [{\citenamefont {Fischer}\ \emph {et~al.}(2016)\citenamefont
  {Fischer}, \citenamefont {Maksymenko},\ and\ \citenamefont
  {Altman}}]{Fischer_2016}%
  \BibitemOpen
  \bibfield  {author} {\bibinfo {author} {\bibfnamefont {M.~H.}\ \bibnamefont
  {Fischer}}, \bibinfo {author} {\bibfnamefont {M.}~\bibnamefont
  {Maksymenko}},\ and\ \bibinfo {author} {\bibfnamefont {E.}~\bibnamefont
  {Altman}},\ }\href
  {https://doi.org/https://doi.org/10.1103/PhysRevLett.116.160401} {\bibfield
  {journal} {\bibinfo  {journal} {Physical Review Letters}\ }\textbf {\bibinfo
  {volume} {116}},\ \bibinfo {pages} {160401} (\bibinfo {year}
  {2016})}\BibitemShut {NoStop}%
\bibitem [{\citenamefont {Levi}\ \emph {et~al.}(2016)\citenamefont {Levi},
  \citenamefont {Heyl}, \citenamefont {Lesanovsky},\ and\ \citenamefont
  {Garrahan}}]{Levi_2016}%
  \BibitemOpen
  \bibfield  {author} {\bibinfo {author} {\bibfnamefont {E.}~\bibnamefont
  {Levi}}, \bibinfo {author} {\bibfnamefont {M.}~\bibnamefont {Heyl}}, \bibinfo
  {author} {\bibfnamefont {I.}~\bibnamefont {Lesanovsky}},\ and\ \bibinfo
  {author} {\bibfnamefont {J.~P.}\ \bibnamefont {Garrahan}},\ }\href
  {https://doi.org/https://doi.org/10.1103/PhysRevLett.116.237203} {\bibfield
  {journal} {\bibinfo  {journal} {Physical Review Letters}\ }\textbf {\bibinfo
  {volume} {116}},\ \bibinfo {pages} {237203} (\bibinfo {year}
  {2016})}\BibitemShut {NoStop}%
\bibitem [{\citenamefont {Medvedyeva}\ \emph {et~al.}(2016)\citenamefont
  {Medvedyeva}, \citenamefont {Prosen},\ and\ \citenamefont
  {{\v{Z}}nidari{\v{c}}}}]{Medvedyeva_2016}%
  \BibitemOpen
  \bibfield  {author} {\bibinfo {author} {\bibfnamefont {M.~V.}\ \bibnamefont
  {Medvedyeva}}, \bibinfo {author} {\bibfnamefont {T.}~\bibnamefont {Prosen}},\
  and\ \bibinfo {author} {\bibfnamefont {M.}~\bibnamefont
  {{\v{Z}}nidari{\v{c}}}},\ }\href
  {https://doi.org/https://doi.org/10.1103/PhysRevB.93.094205} {\bibfield
  {journal} {\bibinfo  {journal} {Physical Review B}\ }\textbf {\bibinfo
  {volume} {93}},\ \bibinfo {pages} {094205} (\bibinfo {year}
  {2016})}\BibitemShut {NoStop}%
\bibitem [{\citenamefont {Everest}\ \emph {et~al.}(2017)\citenamefont
  {Everest}, \citenamefont {Lesanovsky}, \citenamefont {Garrahan},\ and\
  \citenamefont {Levi}}]{Everest_2017}%
  \BibitemOpen
  \bibfield  {author} {\bibinfo {author} {\bibfnamefont {B.}~\bibnamefont
  {Everest}}, \bibinfo {author} {\bibfnamefont {I.}~\bibnamefont {Lesanovsky}},
  \bibinfo {author} {\bibfnamefont {J.~P.}\ \bibnamefont {Garrahan}},\ and\
  \bibinfo {author} {\bibfnamefont {E.}~\bibnamefont {Levi}},\ }\href
  {https://doi.org/https://doi.org/10.1103/PhysRevB.95.024310} {\bibfield
  {journal} {\bibinfo  {journal} {Physical Review B}\ }\textbf {\bibinfo
  {volume} {95}},\ \bibinfo {pages} {024310} (\bibinfo {year}
  {2017})}\BibitemShut {NoStop}%
\bibitem [{\citenamefont {Gopalakrishnan}\ \emph {et~al.}(2017)\citenamefont
  {Gopalakrishnan}, \citenamefont {Islam},\ and\ \citenamefont
  {Knap}}]{Gopalakrishnan_2017}%
  \BibitemOpen
  \bibfield  {author} {\bibinfo {author} {\bibfnamefont {S.}~\bibnamefont
  {Gopalakrishnan}}, \bibinfo {author} {\bibfnamefont {K.~R.}\ \bibnamefont
  {Islam}},\ and\ \bibinfo {author} {\bibfnamefont {M.}~\bibnamefont {Knap}},\
  }\href {https://doi.org/https://doi.org/10.1103/PhysRevLett.119.046601}
  {\bibfield  {journal} {\bibinfo  {journal} {Physical Review Letters}\
  }\textbf {\bibinfo {volume} {119}},\ \bibinfo {pages} {046601} (\bibinfo
  {year} {2017})}\BibitemShut {NoStop}%
\bibitem [{\citenamefont {Lazarides}\ and\ \citenamefont
  {Moessner}(2017)}]{Lazarides_2017}%
  \BibitemOpen
  \bibfield  {author} {\bibinfo {author} {\bibfnamefont {A.}~\bibnamefont
  {Lazarides}}\ and\ \bibinfo {author} {\bibfnamefont {R.}~\bibnamefont
  {Moessner}},\ }\href
  {https://doi.org/https://doi.org/10.1103/PhysRevB.95.195135} {\bibfield
  {journal} {\bibinfo  {journal} {Physical Review B}\ }\textbf {\bibinfo
  {volume} {95}},\ \bibinfo {pages} {195135} (\bibinfo {year}
  {2017})}\BibitemShut {NoStop}%
\bibitem [{\citenamefont {Wu}\ \emph {et~al.}(2019)\citenamefont {Wu},
  \citenamefont {Schnell}, \citenamefont {Tomasi}, \citenamefont {Heyl},\ and\
  \citenamefont {Eckardt}}]{Wu_2019}%
  \BibitemOpen
  \bibfield  {author} {\bibinfo {author} {\bibfnamefont {L.-N.}\ \bibnamefont
  {Wu}}, \bibinfo {author} {\bibfnamefont {A.}~\bibnamefont {Schnell}},
  \bibinfo {author} {\bibfnamefont {G.~D.}\ \bibnamefont {Tomasi}}, \bibinfo
  {author} {\bibfnamefont {M.}~\bibnamefont {Heyl}},\ and\ \bibinfo {author}
  {\bibfnamefont {A.}~\bibnamefont {Eckardt}},\ }\href
  {https://doi.org/https://doi.org/10.1088/1367-2630/ab25a4} {\bibfield
  {journal} {\bibinfo  {journal} {New Journal of Physics}\ }\textbf {\bibinfo
  {volume} {21}},\ \bibinfo {pages} {063026} (\bibinfo {year}
  {2019})}\BibitemShut {NoStop}%
\bibitem [{\citenamefont {Ren}\ \emph {et~al.}(2020)\citenamefont {Ren},
  \citenamefont {Li}, \citenamefont {Li}, \citenamefont {Cai},\ and\
  \citenamefont {Wang}}]{Ren_2020}%
  \BibitemOpen
  \bibfield  {author} {\bibinfo {author} {\bibfnamefont {J.}~\bibnamefont
  {Ren}}, \bibinfo {author} {\bibfnamefont {Q.}~\bibnamefont {Li}}, \bibinfo
  {author} {\bibfnamefont {W.}~\bibnamefont {Li}}, \bibinfo {author}
  {\bibfnamefont {Z.}~\bibnamefont {Cai}},\ and\ \bibinfo {author}
  {\bibfnamefont {X.}~\bibnamefont {Wang}},\ }\href
  {https://doi.org/https://doi.org/10.1103/PhysRevLett.124.130602} {\bibfield
  {journal} {\bibinfo  {journal} {Physical Review Letters}\ }\textbf {\bibinfo
  {volume} {124}},\ \bibinfo {pages} {130602} (\bibinfo {year}
  {2020})}\BibitemShut {NoStop}%
\bibitem [{\citenamefont {Nandkishore}\ and\ \citenamefont
  {Gopalakrishnan}(2016)}]{Nandkishore_2016}%
  \BibitemOpen
  \bibfield  {author} {\bibinfo {author} {\bibfnamefont {R.}~\bibnamefont
  {Nandkishore}}\ and\ \bibinfo {author} {\bibfnamefont {S.}~\bibnamefont
  {Gopalakrishnan}},\ }\href
  {https://doi.org/https://doi.org/10.1002/andp.201600181} {\bibfield
  {journal} {\bibinfo  {journal} {Annalen der Physik}\ }\textbf {\bibinfo
  {volume} {529}},\ \bibinfo {pages} {1600181} (\bibinfo {year}
  {2016})}\BibitemShut {NoStop}%
\bibitem [{\citenamefont {De~Roeck}\ and\ \citenamefont
  {Huveneers}(2017)}]{De_Roeck_2017}%
  \BibitemOpen
  \bibfield  {author} {\bibinfo {author} {\bibfnamefont {W.}~\bibnamefont
  {De~Roeck}}\ and\ \bibinfo {author} {\bibfnamefont {F.}~\bibnamefont
  {Huveneers}},\ }\href
  {https://doi.org/https://doi.org/10.1103/PhysRevB.95.155129} {\bibfield
  {journal} {\bibinfo  {journal} {Physical Review B}\ }\textbf {\bibinfo
  {volume} {95}},\ \bibinfo {pages} {155129} (\bibinfo {year}
  {2017})}\BibitemShut {NoStop}%
\bibitem [{\citenamefont {Luitz}\ \emph {et~al.}(2017)\citenamefont {Luitz},
  \citenamefont {Huveneers},\ and\ \citenamefont {De~Roeck}}]{Luitz_2017}%
  \BibitemOpen
  \bibfield  {author} {\bibinfo {author} {\bibfnamefont {D.~J.}\ \bibnamefont
  {Luitz}}, \bibinfo {author} {\bibfnamefont {F.}~\bibnamefont {Huveneers}},\
  and\ \bibinfo {author} {\bibfnamefont {W.}~\bibnamefont {De~Roeck}},\ }\href
  {https://doi.org/https://doi.org/10.1103/PhysRevLett.119.150602} {\bibfield
  {journal} {\bibinfo  {journal} {Physical Review Letters}\ }\textbf {\bibinfo
  {volume} {119}},\ \bibinfo {pages} {150602} (\bibinfo {year}
  {2017})}\BibitemShut {NoStop}%
\bibitem [{\citenamefont {L{\"u}schen}\ \emph {et~al.}(2017)\citenamefont
  {L{\"u}schen}, \citenamefont {Bordia}, \citenamefont {Hodgman}, \citenamefont
  {Schreiber}, \citenamefont {Sarkar}, \citenamefont {Daley}, \citenamefont
  {Fischer}, \citenamefont {Altman}, \citenamefont {Bloch},\ and\ \citenamefont
  {Schneider}}]{L_schen_2017}%
  \BibitemOpen
  \bibfield  {author} {\bibinfo {author} {\bibfnamefont {H.~P.}\ \bibnamefont
  {L{\"u}schen}}, \bibinfo {author} {\bibfnamefont {P.}~\bibnamefont {Bordia}},
  \bibinfo {author} {\bibfnamefont {S.~S.}\ \bibnamefont {Hodgman}}, \bibinfo
  {author} {\bibfnamefont {M.}~\bibnamefont {Schreiber}}, \bibinfo {author}
  {\bibfnamefont {S.}~\bibnamefont {Sarkar}}, \bibinfo {author} {\bibfnamefont
  {A.~J.}\ \bibnamefont {Daley}}, \bibinfo {author} {\bibfnamefont {M.~H.}\
  \bibnamefont {Fischer}}, \bibinfo {author} {\bibfnamefont {E.}~\bibnamefont
  {Altman}}, \bibinfo {author} {\bibfnamefont {I.}~\bibnamefont {Bloch}},\ and\
  \bibinfo {author} {\bibfnamefont {U.}~\bibnamefont {Schneider}},\ }\href
  {https://doi.org/https://doi.org/10.1103/PhysRevX.7.011034} {\bibfield
  {journal} {\bibinfo  {journal} {Physical Review X}\ }\textbf {\bibinfo
  {volume} {7}},\ \bibinfo {pages} {011034} (\bibinfo {year}
  {2017})}\BibitemShut {NoStop}%
\bibitem [{\citenamefont {Rubio-Abadal}\ \emph {et~al.}(2019)\citenamefont
  {Rubio-Abadal}, \citenamefont {yoon Choi}, \citenamefont {Zeiher},
  \citenamefont {Hollerith}, \citenamefont {Rui}, \citenamefont {Bloch},\ and\
  \citenamefont {Gross}}]{Rubio_Abadal_2019}%
  \BibitemOpen
  \bibfield  {author} {\bibinfo {author} {\bibfnamefont {A.}~\bibnamefont
  {Rubio-Abadal}}, \bibinfo {author} {\bibfnamefont {J.}~\bibnamefont {yoon
  Choi}}, \bibinfo {author} {\bibfnamefont {J.}~\bibnamefont {Zeiher}},
  \bibinfo {author} {\bibfnamefont {S.}~\bibnamefont {Hollerith}}, \bibinfo
  {author} {\bibfnamefont {J.}~\bibnamefont {Rui}}, \bibinfo {author}
  {\bibfnamefont {I.}~\bibnamefont {Bloch}},\ and\ \bibinfo {author}
  {\bibfnamefont {C.}~\bibnamefont {Gross}},\ }\href
  {https://doi.org/https://doi.org/10.1103/PhysRevX.9.041014} {\bibfield
  {journal} {\bibinfo  {journal} {Physical Review X}\ }\textbf {\bibinfo
  {volume} {9}},\ \bibinfo {pages} {041014} (\bibinfo {year}
  {2019})}\BibitemShut {NoStop}%
\bibitem [{\citenamefont {Huse}\ \emph {et~al.}(2015)\citenamefont {Huse},
  \citenamefont {Nandkishore}, \citenamefont {Pietracaprina}, \citenamefont
  {Ros},\ and\ \citenamefont {Scardicchio}}]{Huse_2015}%
  \BibitemOpen
  \bibfield  {author} {\bibinfo {author} {\bibfnamefont {D.~A.}\ \bibnamefont
  {Huse}}, \bibinfo {author} {\bibfnamefont {R.}~\bibnamefont {Nandkishore}},
  \bibinfo {author} {\bibfnamefont {F.}~\bibnamefont {Pietracaprina}}, \bibinfo
  {author} {\bibfnamefont {V.}~\bibnamefont {Ros}},\ and\ \bibinfo {author}
  {\bibfnamefont {A.}~\bibnamefont {Scardicchio}},\ }\href
  {https://doi.org/https://doi.org/10.1103/PhysRevB.92.014203} {\bibfield
  {journal} {\bibinfo  {journal} {Physical Review B}\ }\textbf {\bibinfo
  {volume} {92}},\ \bibinfo {pages} {014203} (\bibinfo {year}
  {2015})}\BibitemShut {NoStop}%
\bibitem [{\citenamefont {Marino}\ and\ \citenamefont
  {Nandkishore}(2018)}]{Marino_2018}%
  \BibitemOpen
  \bibfield  {author} {\bibinfo {author} {\bibfnamefont {J.}~\bibnamefont
  {Marino}}\ and\ \bibinfo {author} {\bibfnamefont {R.~M.}\ \bibnamefont
  {Nandkishore}},\ }\href
  {https://doi.org/https://doi.org/10.1103/PhysRevB.97.054201} {\bibfield
  {journal} {\bibinfo  {journal} {Physical Review B}\ }\textbf {\bibinfo
  {volume} {97}},\ \bibinfo {pages} {054201} (\bibinfo {year}
  {2018})}\BibitemShut {NoStop}%
\bibitem [{\citenamefont {Khemani}\ \emph {et~al.}(2019)\citenamefont
  {Khemani}, \citenamefont {Moessner},\ and\ \citenamefont
  {Sondhi}}]{Khemani_2019}%
  \BibitemOpen
  \bibfield  {author} {\bibinfo {author} {\bibfnamefont {V.}~\bibnamefont
  {Khemani}}, \bibinfo {author} {\bibfnamefont {R.}~\bibnamefont {Moessner}},\
  and\ \bibinfo {author} {\bibfnamefont {S.~L.}\ \bibnamefont {Sondhi}}\ }\href
  {https://doi.org/https://doi.org/10.48550/arXiv.1910.10745}
  {https://doi.org/10.48550/arXiv.1910.10745} (\bibinfo {year}
  {2019})\BibitemShut {NoStop}%
\bibitem [{\citenamefont {Zaletel}\ \emph {et~al.}(2023)\citenamefont
  {Zaletel}, \citenamefont {Lukin}, \citenamefont {Monroe}, \citenamefont
  {Nayak}, \citenamefont {Wilczek},\ and\ \citenamefont {Yao}}]{Zalatel2023}%
  \BibitemOpen
  \bibfield  {author} {\bibinfo {author} {\bibfnamefont {M.~P.}\ \bibnamefont
  {Zaletel}}, \bibinfo {author} {\bibfnamefont {M.}~\bibnamefont {Lukin}},
  \bibinfo {author} {\bibfnamefont {C.}~\bibnamefont {Monroe}}, \bibinfo
  {author} {\bibfnamefont {C.}~\bibnamefont {Nayak}}, \bibinfo {author}
  {\bibfnamefont {F.}~\bibnamefont {Wilczek}},\ and\ \bibinfo {author}
  {\bibfnamefont {N.~Y.}\ \bibnamefont {Yao}},\ }\href
  {https://doi.org/https://doi.org/10.1103/RevModPhys.95.031001} {\bibfield
  {journal} {\bibinfo  {journal} {Reviews of Modern Physics}\ }\textbf
  {\bibinfo {volume} {95}},\ \bibinfo {pages} {031001} (\bibinfo {year}
  {2023})}\BibitemShut {NoStop}%
\bibitem [{\citenamefont {Sieberer}\ \emph {et~al.}(2023)\citenamefont
  {Sieberer}, \citenamefont {Buchhold}, \citenamefont {Marino},\ and\
  \citenamefont {Diehl}}]{Sieberer2023}%
  \BibitemOpen
  \bibfield  {author} {\bibinfo {author} {\bibfnamefont {L.~M.}\ \bibnamefont
  {Sieberer}}, \bibinfo {author} {\bibfnamefont {M.}~\bibnamefont {Buchhold}},
  \bibinfo {author} {\bibfnamefont {J.}~\bibnamefont {Marino}},\ and\ \bibinfo
  {author} {\bibfnamefont {S.}~\bibnamefont {Diehl}}\ }\href
  {https://doi.org/https://doi.org/10.48550/arXiv.2312.03073}
  {https://doi.org/10.48550/arXiv.2312.03073} (\bibinfo {year}
  {2023})\BibitemShut {NoStop}%
\bibitem [{\citenamefont {Ros}\ and\ \citenamefont {M\"uller}()}]{Ros_2017}%
  \BibitemOpen
  \bibfield  {author} {\bibinfo {author} {\bibfnamefont {V.}~\bibnamefont
  {Ros}}\ and\ \bibinfo {author} {\bibfnamefont {M.}~\bibnamefont {M\"uller}},\
  }\href {https://doi.org/http://dx.doi.org/10.1103/PhysRevLett.118.237202}
  {\bibfield  {journal} {\bibinfo  {journal} {Physical Review Letters}\
  }\textbf {\bibinfo {volume} {118}},\ \bibinfo {pages} {237202}}\BibitemShut
  {NoStop}%
\bibitem [{\citenamefont {Lenar{\v{c}}i{\v{c}}}\ \emph
  {et~al.}(2018)\citenamefont {Lenar{\v{c}}i{\v{c}}}, \citenamefont {Altman},\
  and\ \citenamefont {Rosch}}]{Lenar_i__2018}%
  \BibitemOpen
  \bibfield  {author} {\bibinfo {author} {\bibfnamefont {Z.}~\bibnamefont
  {Lenar{\v{c}}i{\v{c}}}}, \bibinfo {author} {\bibfnamefont {E.}~\bibnamefont
  {Altman}},\ and\ \bibinfo {author} {\bibfnamefont {A.}~\bibnamefont
  {Rosch}},\ }\href
  {https://doi.org/https://doi.org/10.1103/PhysRevLett.121.267603} {\bibfield
  {journal} {\bibinfo  {journal} {Physical Review Letters}\ }\textbf {\bibinfo
  {volume} {121}},\ \bibinfo {pages} {267603} (\bibinfo {year}
  {2018})}\BibitemShut {NoStop}%
\bibitem [{\citenamefont {Lenar{\v{c}}i{\v{c}}}\ \emph
  {et~al.}(2020)\citenamefont {Lenar{\v{c}}i{\v{c}}}, \citenamefont {Alberton},
  \citenamefont {Rosch},\ and\ \citenamefont {Altman}}]{Lenar_i__2020}%
  \BibitemOpen
  \bibfield  {author} {\bibinfo {author} {\bibfnamefont {Z.}~\bibnamefont
  {Lenar{\v{c}}i{\v{c}}}}, \bibinfo {author} {\bibfnamefont {O.}~\bibnamefont
  {Alberton}}, \bibinfo {author} {\bibfnamefont {A.}~\bibnamefont {Rosch}},\
  and\ \bibinfo {author} {\bibfnamefont {E.}~\bibnamefont {Altman}},\ }\href
  {https://doi.org/https://doi.org/10.1103/PhysRevLett.125.116601} {\bibfield
  {journal} {\bibinfo  {journal} {Physical Review Letters}\ }\textbf {\bibinfo
  {volume} {125}},\ \bibinfo {pages} {116601} (\bibinfo {year}
  {2020})}\BibitemShut {NoStop}%
\bibitem [{\citenamefont {Takayoshi}\ \emph
  {et~al.}(2014{\natexlab{a}})\citenamefont {Takayoshi}, \citenamefont {Aoki},\
  and\ \citenamefont {Oka}}]{Takayoshi_2014}%
  \BibitemOpen
  \bibfield  {author} {\bibinfo {author} {\bibfnamefont {S.}~\bibnamefont
  {Takayoshi}}, \bibinfo {author} {\bibfnamefont {H.}~\bibnamefont {Aoki}},\
  and\ \bibinfo {author} {\bibfnamefont {T.}~\bibnamefont {Oka}},\ }\href
  {https://doi.org/http://dx.doi.org/10.1103/PhysRevB.90.085150} {\bibfield
  {journal} {\bibinfo  {journal} {Physical Review B}\ }\textbf {\bibinfo
  {volume} {90}},\ \bibinfo {pages} {085150} (\bibinfo {year}
  {2014}{\natexlab{a}})}\BibitemShut {NoStop}%
\bibitem [{\citenamefont {Takayoshi}\ \emph
  {et~al.}(2014{\natexlab{b}})\citenamefont {Takayoshi}, \citenamefont {Sato},\
  and\ \citenamefont {Oka}}]{Takayoshi_2014_2}%
  \BibitemOpen
  \bibfield  {author} {\bibinfo {author} {\bibfnamefont {S.}~\bibnamefont
  {Takayoshi}}, \bibinfo {author} {\bibfnamefont {M.}~\bibnamefont {Sato}},\
  and\ \bibinfo {author} {\bibfnamefont {T.}~\bibnamefont {Oka}},\ }\href
  {https://doi.org/http://dx.doi.org/10.1103/PhysRevB.90.214413} {\bibfield
  {journal} {\bibinfo  {journal} {Physical Review B}\ }\textbf {\bibinfo
  {volume} {90}},\ \bibinfo {pages} {214413} (\bibinfo {year}
  {2014}{\natexlab{b}})}\BibitemShut {NoStop}%
\bibitem [{\citenamefont {Herbrych}\ and\ \citenamefont
  {Zotos}(2016)}]{Herbrych_2016}%
  \BibitemOpen
  \bibfield  {author} {\bibinfo {author} {\bibfnamefont {J.}~\bibnamefont
  {Herbrych}}\ and\ \bibinfo {author} {\bibfnamefont {X.}~\bibnamefont
  {Zotos}},\ }\href
  {https://doi.org/http://dx.doi.org/10.1103/PhysRevB.93.134412} {\bibfield
  {journal} {\bibinfo  {journal} {Physical Review B}\ }\textbf {\bibinfo
  {volume} {93}},\ \bibinfo {pages} {134412} (\bibinfo {year}
  {2016})}\BibitemShut {NoStop}%
\bibitem [{\citenamefont {Richter}\ \emph {et~al.}(2018)\citenamefont
  {Richter}, \citenamefont {Herbrych},\ and\ \citenamefont
  {Steinigeweg}}]{Richter_2018}%
  \BibitemOpen
  \bibfield  {author} {\bibinfo {author} {\bibfnamefont {J.}~\bibnamefont
  {Richter}}, \bibinfo {author} {\bibfnamefont {J.}~\bibnamefont {Herbrych}},\
  and\ \bibinfo {author} {\bibfnamefont {R.}~\bibnamefont {Steinigeweg}},\
  }\href {https://doi.org/https://doi.org/10.1103/PhysRevB.98.134302}
  {\bibfield  {journal} {\bibinfo  {journal} {Physical Review B}\ }\textbf
  {\bibinfo {volume} {98}},\ \bibinfo {pages} {134302} (\bibinfo {year}
  {2018})}\BibitemShut {NoStop}%
\bibitem [{\citenamefont {Richter}\ \emph
  {et~al.}(2019{\natexlab{a}})\citenamefont {Richter}, \citenamefont {Gemmer},\
  and\ \citenamefont {Steinigeweg}}]{Richter_2019}%
  \BibitemOpen
  \bibfield  {author} {\bibinfo {author} {\bibfnamefont {J.}~\bibnamefont
  {Richter}}, \bibinfo {author} {\bibfnamefont {J.}~\bibnamefont {Gemmer}},\
  and\ \bibinfo {author} {\bibfnamefont {R.}~\bibnamefont {Steinigeweg}},\
  }\href {https://doi.org/https://doi.org/10.1103/PhysRevE.99.050104}
  {\bibfield  {journal} {\bibinfo  {journal} {Physical Review E}\ }\textbf
  {\bibinfo {volume} {99}},\ \bibinfo {pages} {050104} (\bibinfo {year}
  {2019}{\natexlab{a}})}\BibitemShut {NoStop}%
\bibitem [{\citenamefont {Richter}\ and\ \citenamefont
  {Steinigeweg}(2019{\natexlab{a}})}]{Richter_2019_2}%
  \BibitemOpen
  \bibfield  {author} {\bibinfo {author} {\bibfnamefont {J.}~\bibnamefont
  {Richter}}\ and\ \bibinfo {author} {\bibfnamefont {R.}~\bibnamefont
  {Steinigeweg}},\ }\href
  {https://doi.org/https://doi.org/10.1103/PhysRevE.99.012114} {\bibfield
  {journal} {\bibinfo  {journal} {Physical Review E}\ }\textbf {\bibinfo
  {volume} {99}},\ \bibinfo {pages} {012114} (\bibinfo {year}
  {2019}{\natexlab{a}})}\BibitemShut {NoStop}%
\bibitem [{\citenamefont {Breuer}\ and\ \citenamefont
  {Petruccione}(2007)}]{Breuer_2007}%
  \BibitemOpen
  \bibfield  {author} {\bibinfo {author} {\bibfnamefont {H.-P.}\ \bibnamefont
  {Breuer}}\ and\ \bibinfo {author} {\bibfnamefont {F.}~\bibnamefont
  {Petruccione}},\ }\href
  {https://doi.org/http://dx.doi.org/10.1093/acprof:oso/9780199213900.001.0001}
  {\emph {\bibinfo {title} {The Theory of Open Quantum Systems}}}\ (\bibinfo
  {publisher} {Oxford University {PressOxford}},\ \bibinfo {year}
  {2007})\BibitemShut {NoStop}%
\bibitem [{\citenamefont {{\v{Z}}nidari{\v{c}}}\ \emph
  {et~al.}(2016)\citenamefont {{\v{Z}}nidari{\v{c}}}, \citenamefont
  {Mendoza-Arenas}, \citenamefont {Clark},\ and\ \citenamefont
  {Goold}}]{_nidari__2016}%
  \BibitemOpen
  \bibfield  {author} {\bibinfo {author} {\bibfnamefont {M.}~\bibnamefont
  {{\v{Z}}nidari{\v{c}}}}, \bibinfo {author} {\bibfnamefont {J.~J.}\
  \bibnamefont {Mendoza-Arenas}}, \bibinfo {author} {\bibfnamefont {S.~R.}\
  \bibnamefont {Clark}},\ and\ \bibinfo {author} {\bibfnamefont
  {J.}~\bibnamefont {Goold}},\ }\href
  {https://doi.org/https://doi.org/10.1002/andp.201600298} {\bibfield
  {journal} {\bibinfo  {journal} {Annalen der Physik}\ }\textbf {\bibinfo
  {volume} {529}},\ \bibinfo {pages} {1600298} (\bibinfo {year}
  {2016})}\BibitemShut {NoStop}%
\bibitem [{\citenamefont {Fehske}\ \emph {et~al.}(2009)\citenamefont {Fehske},
  \citenamefont {Schleede}, \citenamefont {Schubert}, \citenamefont {Wellein},
  \citenamefont {Filinov},\ and\ \citenamefont {Bishop}}]{Fehske_2009}%
  \BibitemOpen
  \bibfield  {author} {\bibinfo {author} {\bibfnamefont {H.}~\bibnamefont
  {Fehske}}, \bibinfo {author} {\bibfnamefont {J.}~\bibnamefont {Schleede}},
  \bibinfo {author} {\bibfnamefont {G.}~\bibnamefont {Schubert}}, \bibinfo
  {author} {\bibfnamefont {G.}~\bibnamefont {Wellein}}, \bibinfo {author}
  {\bibfnamefont {V.~S.}\ \bibnamefont {Filinov}},\ and\ \bibinfo {author}
  {\bibfnamefont {A.~R.}\ \bibnamefont {Bishop}},\ }\href
  {https://doi.org/10.1016/j.physleta.2009.04.022} {\bibfield  {journal}
  {\bibinfo  {journal} {Phys. Lett. A}\ }\textbf {\bibinfo {volume} {373}},\
  \bibinfo {pages} {2182} (\bibinfo {year} {2009})}\BibitemShut {NoStop}%
\bibitem [{\citenamefont {Heitmann}\ \emph {et~al.}(2020)\citenamefont
  {Heitmann}, \citenamefont {Richter}, \citenamefont {Schubert},\ and\
  \citenamefont {Steinigeweg}}]{Heitmann_2020}%
  \BibitemOpen
  \bibfield  {author} {\bibinfo {author} {\bibfnamefont {T.}~\bibnamefont
  {Heitmann}}, \bibinfo {author} {\bibfnamefont {J.}~\bibnamefont {Richter}},
  \bibinfo {author} {\bibfnamefont {D.}~\bibnamefont {Schubert}},\ and\
  \bibinfo {author} {\bibfnamefont {R.}~\bibnamefont {Steinigeweg}},\ }\href
  {https://doi.org/https://doi.org/10.1515/zna-2020-0010} {\bibfield  {journal}
  {\bibinfo  {journal} {Zeitschrift f{\"u}r Naturforschung A}\ }\textbf
  {\bibinfo {volume} {75}},\ \bibinfo {pages} {421} (\bibinfo {year}
  {2020})}\BibitemShut {NoStop}%
\bibitem [{\citenamefont {Jin}\ \emph {et~al.}(2021)\citenamefont {Jin},
  \citenamefont {Willsch}, \citenamefont {Willsch}, \citenamefont {Lagemann},
  \citenamefont {Michielsen},\ and\ \citenamefont {Raedt}}]{Jin_2021}%
  \BibitemOpen
  \bibfield  {author} {\bibinfo {author} {\bibfnamefont {F.}~\bibnamefont
  {Jin}}, \bibinfo {author} {\bibfnamefont {D.}~\bibnamefont {Willsch}},
  \bibinfo {author} {\bibfnamefont {M.}~\bibnamefont {Willsch}}, \bibinfo
  {author} {\bibfnamefont {H.}~\bibnamefont {Lagemann}}, \bibinfo {author}
  {\bibfnamefont {K.}~\bibnamefont {Michielsen}},\ and\ \bibinfo {author}
  {\bibfnamefont {H.~D.}\ \bibnamefont {Raedt}},\ }\href
  {https://doi.org/https://doi.org/10.7566/JPSJ.90.012001} {\bibfield
  {journal} {\bibinfo  {journal} {Journal of the Physical Society of Japan}\
  }\textbf {\bibinfo {volume} {90}},\ \bibinfo {pages} {012001} (\bibinfo
  {year} {2021})}\BibitemShut {NoStop}%
\bibitem [{\citenamefont {Hams}\ and\ \citenamefont {{De
  Raedt}}(2000)}]{Hams2000}%
  \BibitemOpen
  \bibfield  {author} {\bibinfo {author} {\bibfnamefont {A.}~\bibnamefont
  {Hams}}\ and\ \bibinfo {author} {\bibfnamefont {H.}~\bibnamefont {{De
  Raedt}}},\ }\href {https://doi.org/10.1103/PhysRevE.62.4365} {\bibfield
  {journal} {\bibinfo  {journal} {Phys. Rev. E}\ }\textbf {\bibinfo {volume}
  {62}},\ \bibinfo {pages} {4365} (\bibinfo {year} {2000})}\BibitemShut
  {NoStop}%
\bibitem [{\citenamefont {Sugiura}\ and\ \citenamefont
  {Shimizu}(2013)}]{Sugiura2013}%
  \BibitemOpen
  \bibfield  {author} {\bibinfo {author} {\bibfnamefont {S.}~\bibnamefont
  {Sugiura}}\ and\ \bibinfo {author} {\bibfnamefont {A.}~\bibnamefont
  {Shimizu}},\ }\href {https://doi.org/10.1103/PhysRevLett.111.010401}
  {\bibfield  {journal} {\bibinfo  {journal} {Phys. Rev. Lett.}\ }\textbf
  {\bibinfo {volume} {111}},\ \bibinfo {pages} {010401} (\bibinfo {year}
  {2013})}\BibitemShut {NoStop}%
\bibitem [{\citenamefont {Dalibard}\ \emph {et~al.}(1992)\citenamefont
  {Dalibard}, \citenamefont {Castin},\ and\ \citenamefont
  {M{\o}lmer}}]{Dalibard_1992}%
  \BibitemOpen
  \bibfield  {author} {\bibinfo {author} {\bibfnamefont {J.}~\bibnamefont
  {Dalibard}}, \bibinfo {author} {\bibfnamefont {Y.}~\bibnamefont {Castin}},\
  and\ \bibinfo {author} {\bibfnamefont {K.}~\bibnamefont {M{\o}lmer}},\ }\href
  {https://doi.org/http://dx.doi.org/ 10.1103/PhysRevLett.68.580} {\bibfield
  {journal} {\bibinfo  {journal} {Physical Review Letters}\ }\textbf {\bibinfo
  {volume} {68}},\ \bibinfo {pages} {580} (\bibinfo {year} {1992})}\BibitemShut
  {NoStop}%
\bibitem [{\citenamefont {Heitmann}\ \emph
  {et~al.}(2023{\natexlab{a}})\citenamefont {Heitmann}, \citenamefont
  {Richter}, \citenamefont {Herbrych}, \citenamefont {Gemmer},\ and\
  \citenamefont {Steinigeweg}}]{Heitmann_2022}%
  \BibitemOpen
  \bibfield  {author} {\bibinfo {author} {\bibfnamefont {T.}~\bibnamefont
  {Heitmann}}, \bibinfo {author} {\bibfnamefont {J.}~\bibnamefont {Richter}},
  \bibinfo {author} {\bibfnamefont {J.}~\bibnamefont {Herbrych}}, \bibinfo
  {author} {\bibfnamefont {J.}~\bibnamefont {Gemmer}},\ and\ \bibinfo {author}
  {\bibfnamefont {R.}~\bibnamefont {Steinigeweg}},\ }\href
  {https://doi.org/https://doi.org/10.1103/PhysRevE.108.024102} {\bibfield
  {journal} {\bibinfo  {journal} {Physical Review E}\ }\textbf {\bibinfo
  {volume} {108}},\ \bibinfo {pages} {024102} (\bibinfo {year}
  {2023}{\natexlab{a}})}\BibitemShut {NoStop}%
\bibitem [{\citenamefont {Heitmann}\ \emph
  {et~al.}(2023{\natexlab{b}})\citenamefont {Heitmann}, \citenamefont
  {Richter}, \citenamefont {Jin}, \citenamefont {Nandy}, \citenamefont
  {Lenar{\v{c}}i{\v{c}}}, \citenamefont {Herbrych}, \citenamefont {Michielsen},
  \citenamefont {De~Raedt}, \citenamefont {Gemmer},\ and\ \citenamefont
  {Steinigeweg}}]{Heitmann_2023}%
  \BibitemOpen
  \bibfield  {author} {\bibinfo {author} {\bibfnamefont {T.}~\bibnamefont
  {Heitmann}}, \bibinfo {author} {\bibfnamefont {J.}~\bibnamefont {Richter}},
  \bibinfo {author} {\bibfnamefont {F.}~\bibnamefont {Jin}}, \bibinfo {author}
  {\bibfnamefont {S.}~\bibnamefont {Nandy}}, \bibinfo {author} {\bibfnamefont
  {Z.}~\bibnamefont {Lenar{\v{c}}i{\v{c}}}}, \bibinfo {author} {\bibfnamefont
  {J.}~\bibnamefont {Herbrych}}, \bibinfo {author} {\bibfnamefont
  {K.}~\bibnamefont {Michielsen}}, \bibinfo {author} {\bibfnamefont
  {H.}~\bibnamefont {De~Raedt}}, \bibinfo {author} {\bibfnamefont
  {J.}~\bibnamefont {Gemmer}},\ and\ \bibinfo {author} {\bibfnamefont
  {R.}~\bibnamefont {Steinigeweg}},\ }\href
  {https://doi.org/https://doi.org/10.1103/PhysRevB.108.L201119} {\bibfield
  {journal} {\bibinfo  {journal} {Physical Review B}\ }\textbf {\bibinfo
  {volume} {108}},\ \bibinfo {pages} {l201119} (\bibinfo {year}
  {2023}{\natexlab{b}})}\BibitemShut {NoStop}%
\bibitem [{\citenamefont {Dabelow}\ and\ \citenamefont
  {Reimann}(2024)}]{Dabelow_2024}%
  \BibitemOpen
  \bibfield  {author} {\bibinfo {author} {\bibfnamefont {L.}~\bibnamefont
  {Dabelow}}\ and\ \bibinfo {author} {\bibfnamefont {P.}~\bibnamefont
  {Reimann}},\ }\href
  {https://doi.org/https://doi.org/10.1038/s41467-023-44487-2} {\bibfield
  {journal} {\bibinfo  {journal} {Nature Communications}\ }\textbf {\bibinfo
  {volume} {15}},\ \bibinfo {pages} {294} (\bibinfo {year} {2024})}\BibitemShut
  {NoStop}%
\bibitem [{\citenamefont {Richter}\ \emph
  {et~al.}(2019{\natexlab{b}})\citenamefont {Richter}, \citenamefont {Lamann},
  \citenamefont {Bartsch}, \citenamefont {Steinigeweg},\ and\ \citenamefont
  {Gemmer}}]{Richter_2019_3}%
  \BibitemOpen
  \bibfield  {author} {\bibinfo {author} {\bibfnamefont {J.}~\bibnamefont
  {Richter}}, \bibinfo {author} {\bibfnamefont {M.~H.}\ \bibnamefont {Lamann}},
  \bibinfo {author} {\bibfnamefont {C.}~\bibnamefont {Bartsch}}, \bibinfo
  {author} {\bibfnamefont {R.}~\bibnamefont {Steinigeweg}},\ and\ \bibinfo
  {author} {\bibfnamefont {J.}~\bibnamefont {Gemmer}},\ }\href
  {https://doi.org/https://doi.org/10.1103/PhysRevE.100.032124} {\bibfield
  {journal} {\bibinfo  {journal} {Physical Review E}\ }\textbf {\bibinfo
  {volume} {100}},\ \bibinfo {pages} {032124} (\bibinfo {year}
  {2019}{\natexlab{b}})}\BibitemShut {NoStop}%
\bibitem [{\citenamefont {Abanin}\ \emph {et~al.}(2017)\citenamefont {Abanin},
  \citenamefont {De~Roeck}, \citenamefont {Ho},\ and\ \citenamefont
  {Huveneers}}]{Abanin_2017}%
  \BibitemOpen
  \bibfield  {author} {\bibinfo {author} {\bibfnamefont {D.}~\bibnamefont
  {Abanin}}, \bibinfo {author} {\bibfnamefont {W.}~\bibnamefont {De~Roeck}},
  \bibinfo {author} {\bibfnamefont {W.~W.}\ \bibnamefont {Ho}},\ and\ \bibinfo
  {author} {\bibfnamefont {F.}~\bibnamefont {Huveneers}},\ }\href
  {https://doi.org/https://doi.org/10.1007/s00220-017-2930-x} {\bibfield
  {journal} {\bibinfo  {journal} {Communications in Mathematical Physics}\
  }\textbf {\bibinfo {volume} {354}},\ \bibinfo {pages} {809} (\bibinfo {year}
  {2017})}\BibitemShut {NoStop}%
\bibitem [{\citenamefont {Kuwahara}\ \emph {et~al.}(2016)\citenamefont
  {Kuwahara}, \citenamefont {Mori},\ and\ \citenamefont
  {Saito}}]{Kuwahara_2016}%
  \BibitemOpen
  \bibfield  {author} {\bibinfo {author} {\bibfnamefont {T.}~\bibnamefont
  {Kuwahara}}, \bibinfo {author} {\bibfnamefont {T.}~\bibnamefont {Mori}},\
  and\ \bibinfo {author} {\bibfnamefont {K.}~\bibnamefont {Saito}},\ }\href
  {https://doi.org/https://doi.org/10.1016/j.aop.2016.01.012} {\bibfield
  {journal} {\bibinfo  {journal} {Annals of Physics}\ }\textbf {\bibinfo
  {volume} {367}},\ \bibinfo {pages} {96} (\bibinfo {year} {2016})}\BibitemShut
  {NoStop}%
\bibitem [{\citenamefont {Khemani}\ \emph {et~al.}(2016)\citenamefont
  {Khemani}, \citenamefont {Lazarides}, \citenamefont {Moessner},\ and\
  \citenamefont {Sondhi}}]{Khemani_2016}%
  \BibitemOpen
  \bibfield  {author} {\bibinfo {author} {\bibfnamefont {V.}~\bibnamefont
  {Khemani}}, \bibinfo {author} {\bibfnamefont {A.}~\bibnamefont {Lazarides}},
  \bibinfo {author} {\bibfnamefont {R.}~\bibnamefont {Moessner}},\ and\
  \bibinfo {author} {\bibfnamefont {S.}~\bibnamefont {Sondhi}},\ }\href
  {https://doi.org/https://doi.org/10.1103/PhysRevLett.116.250401} {\bibfield
  {journal} {\bibinfo  {journal} {Physical Review Letters}\ }\textbf {\bibinfo
  {volume} {116}},\ \bibinfo {pages} {250401} (\bibinfo {year}
  {2016})}\BibitemShut {NoStop}%
\bibitem [{\citenamefont {O'Sullivan}\ \emph {et~al.}(2020)\citenamefont
  {O'Sullivan}, \citenamefont {Lunt}, \citenamefont {Zollitsch}, \citenamefont
  {Thewalt}, \citenamefont {Morton},\ and\ \citenamefont
  {Pal}}]{O_Sullivan_2020}%
  \BibitemOpen
  \bibfield  {author} {\bibinfo {author} {\bibfnamefont {J.}~\bibnamefont
  {O'Sullivan}}, \bibinfo {author} {\bibfnamefont {O.}~\bibnamefont {Lunt}},
  \bibinfo {author} {\bibfnamefont {C.~W.}\ \bibnamefont {Zollitsch}}, \bibinfo
  {author} {\bibfnamefont {M.~L.~W.}\ \bibnamefont {Thewalt}}, \bibinfo
  {author} {\bibfnamefont {J.~J.~L.}\ \bibnamefont {Morton}},\ and\ \bibinfo
  {author} {\bibfnamefont {A.}~\bibnamefont {Pal}},\ }\href
  {https://doi.org/https://doi.org/10.1088/1367-2630/ab9fbe} {\bibfield
  {journal} {\bibinfo  {journal} {New Journal of Physics}\ }\textbf {\bibinfo
  {volume} {22}},\ \bibinfo {pages} {085001} (\bibinfo {year}
  {2020})}\BibitemShut {NoStop}%
\bibitem [{\citenamefont {Luitz}\ \emph {et~al.}(2020)\citenamefont {Luitz},
  \citenamefont {Moessner}, \citenamefont {Sondhi},\ and\ \citenamefont
  {Khemani}}]{Luitz_2020}%
  \BibitemOpen
  \bibfield  {author} {\bibinfo {author} {\bibfnamefont {D.~J.}\ \bibnamefont
  {Luitz}}, \bibinfo {author} {\bibfnamefont {R.}~\bibnamefont {Moessner}},
  \bibinfo {author} {\bibfnamefont {S.~L.}\ \bibnamefont {Sondhi}},\ and\
  \bibinfo {author} {\bibfnamefont {V.}~\bibnamefont {Khemani}},\ }\href
  {https://doi.org/https://doi.org/10.1103/PhysRevX.10.021046} {\bibfield
  {journal} {\bibinfo  {journal} {Physical Review X}\ }\textbf {\bibinfo
  {volume} {10}},\ \bibinfo {pages} {021046} (\bibinfo {year}
  {2020})}\BibitemShut {NoStop}%
\bibitem [{\citenamefont {Elsayed}\ and\ \citenamefont
  {Fine}(2013)}]{Elsayed2013}%
  \BibitemOpen
  \bibfield  {author} {\bibinfo {author} {\bibfnamefont {T.~A.}\ \bibnamefont
  {Elsayed}}\ and\ \bibinfo {author} {\bibfnamefont {B.~V.}\ \bibnamefont
  {Fine}},\ }\href {https://doi.org/10.1103/PhysRevLett.110.070404} {\bibfield
  {journal} {\bibinfo  {journal} {Phys. Rev. Lett.}\ }\textbf {\bibinfo
  {volume} {110}},\ \bibinfo {pages} {070404} (\bibinfo {year}
  {2013})}\BibitemShut {NoStop}%
\bibitem [{\citenamefont {Iitaka}\ and\ \citenamefont
  {Ebisuzaki}(2003)}]{Iitaka2003}%
  \BibitemOpen
  \bibfield  {author} {\bibinfo {author} {\bibfnamefont {T.}~\bibnamefont
  {Iitaka}}\ and\ \bibinfo {author} {\bibfnamefont {T.}~\bibnamefont
  {Ebisuzaki}},\ }\href {https://doi.org/10.1103/PhysRevLett.90.047203}
  {\bibfield  {journal} {\bibinfo  {journal} {Phys. Rev. Lett.}\ }\textbf
  {\bibinfo {volume} {90}},\ \bibinfo {pages} {047203} (\bibinfo {year}
  {2003})}\BibitemShut {NoStop}%
\bibitem [{\citenamefont {Steinigeweg}\ \emph {et~al.}(2016)\citenamefont
  {Steinigeweg}, \citenamefont {Herbrych}, \citenamefont {Pollmann},\ and\
  \citenamefont {Brenig}}]{Steinigeweg_2016}%
  \BibitemOpen
  \bibfield  {author} {\bibinfo {author} {\bibfnamefont {R.}~\bibnamefont
  {Steinigeweg}}, \bibinfo {author} {\bibfnamefont {J.}~\bibnamefont
  {Herbrych}}, \bibinfo {author} {\bibfnamefont {F.}~\bibnamefont {Pollmann}},\
  and\ \bibinfo {author} {\bibfnamefont {W.}~\bibnamefont {Brenig}},\ }\href
  {https://doi.org/10.1103/physrevb.94.180401} {\bibfield  {journal} {\bibinfo
  {journal} {Phys. Rev. B}\ }\textbf {\bibinfo {volume} {94}},\ \bibinfo
  {pages} {180401} (\bibinfo {year} {2016})}\BibitemShut {NoStop}%
\bibitem [{\citenamefont {Richter}\ and\ \citenamefont
  {Steinigeweg}(2019{\natexlab{b}})}]{Richter2019b}%
  \BibitemOpen
  \bibfield  {author} {\bibinfo {author} {\bibfnamefont {J.}~\bibnamefont
  {Richter}}\ and\ \bibinfo {author} {\bibfnamefont {R.}~\bibnamefont
  {Steinigeweg}},\ }\href {https://doi.org/10.1103/PhysRevB.99.094419}
  {\bibfield  {journal} {\bibinfo  {journal} {Phys. Rev. B}\ }\textbf {\bibinfo
  {volume} {99}},\ \bibinfo {pages} {094419} (\bibinfo {year}
  {2019}{\natexlab{b}})}\BibitemShut {NoStop}%
\bibitem [{\citenamefont {Zwolak}\ and\ \citenamefont
  {Vidal}(2004)}]{Zwolak_2004}%
  \BibitemOpen
  \bibfield  {author} {\bibinfo {author} {\bibfnamefont {M.}~\bibnamefont
  {Zwolak}}\ and\ \bibinfo {author} {\bibfnamefont {G.}~\bibnamefont {Vidal}},\
  }\href {https://doi.org/10.1103/PhysRevLett.93.207205} {\bibfield  {journal}
  {\bibinfo  {journal} {Physical Review Letters}\ }\textbf {\bibinfo {volume}
  {93}},\ \bibinfo {pages} {207205} (\bibinfo {year} {2004})}\BibitemShut
  {NoStop}%
\bibitem [{\citenamefont {Verstraete}\ \emph {et~al.}(2004)\citenamefont
  {Verstraete}, \citenamefont {Garc{\'i}a-Ripoll},\ and\ \citenamefont
  {Cirac}}]{Verstraete_2004}%
  \BibitemOpen
  \bibfield  {author} {\bibinfo {author} {\bibfnamefont {F.}~\bibnamefont
  {Verstraete}}, \bibinfo {author} {\bibfnamefont {J.~J.}\ \bibnamefont
  {Garc{\'i}a-Ripoll}},\ and\ \bibinfo {author} {\bibfnamefont {J.~I.}\
  \bibnamefont {Cirac}},\ }\href
  {https://doi.org/10.1103/PhysRevLett.93.207204} {\bibfield  {journal}
  {\bibinfo  {journal} {Physical Review Letters}\ }\textbf {\bibinfo {volume}
  {93}},\ \bibinfo {pages} {207204} (\bibinfo {year} {2004})}\BibitemShut
  {NoStop}%
\bibitem [{\citenamefont {Lezama}\ \emph {et~al.}(2019)\citenamefont {Lezama},
  \citenamefont {Bera},\ and\ \citenamefont {Bardarson}}]{Lezama_2019}%
  \BibitemOpen
  \bibfield  {author} {\bibinfo {author} {\bibfnamefont {T.~L.~M.}\
  \bibnamefont {Lezama}}, \bibinfo {author} {\bibfnamefont {S.}~\bibnamefont
  {Bera}},\ and\ \bibinfo {author} {\bibfnamefont {J.~H.}\ \bibnamefont
  {Bardarson}},\ }\href
  {https://doi.org/https://doi.org/10.1103/PhysRevB.99.161106} {\bibfield
  {journal} {\bibinfo  {journal} {Physical Review B}\ }\textbf {\bibinfo
  {volume} {99}},\ \bibinfo {pages} {161106} (\bibinfo {year}
  {2019})}\BibitemShut {NoStop}%
\bibitem [{\citenamefont {Crowley}\ and\ \citenamefont
  {Chandran}(2022)}]{Crowley_2022}%
  \BibitemOpen
  \bibfield  {author} {\bibinfo {author} {\bibfnamefont {P.}~\bibnamefont
  {Crowley}}\ and\ \bibinfo {author} {\bibfnamefont {A.}~\bibnamefont
  {Chandran}},\ }\href {https://doi.org/10.21468/SciPostPhys.12.6.201}
  {\bibfield  {journal} {\bibinfo  {journal} {SciPost Physics}\ }\textbf
  {\bibinfo {volume} {12}},\ \bibinfo {pages} {201} (\bibinfo {year}
  {2022})}\BibitemShut {NoStop}%
\bibitem [{\citenamefont {Long}\ \emph {et~al.}(2023)\citenamefont {Long},
  \citenamefont {Crowley}, \citenamefont {Khemani},\ and\ \citenamefont
  {Chandran}}]{Long2022}%
  \BibitemOpen
  \bibfield  {author} {\bibinfo {author} {\bibfnamefont {D.~M.}\ \bibnamefont
  {Long}}, \bibinfo {author} {\bibfnamefont {P.~J.}\ \bibnamefont {Crowley}},
  \bibinfo {author} {\bibfnamefont {V.}~\bibnamefont {Khemani}},\ and\ \bibinfo
  {author} {\bibfnamefont {A.}~\bibnamefont {Chandran}},\ }\href
  {https://doi.org/https://doi.org/10.1103/PhysRevLett.131.106301} {\bibfield
  {journal} {\bibinfo  {journal} {Physical Review Letters}\ }\textbf {\bibinfo
  {volume} {131}},\ \bibinfo {pages} {106301} (\bibinfo {year}
  {2023})}\BibitemShut {NoStop}%
\bibitem [{\citenamefont {Mahoney}\ and\ \citenamefont
  {Richter}(2024)}]{Mahoney2024}%
  \BibitemOpen
  \bibfield  {author} {\bibinfo {author} {\bibfnamefont {D.~E.}\ \bibnamefont
  {Mahoney}}\ and\ \bibinfo {author} {\bibfnamefont {J.}~\bibnamefont
  {Richter}}\ }\href
  {https://doi.org/https://doi.org/10.48550/arXiv.2403.01681}
  {https://doi.org/10.48550/arXiv.2403.01681} (\bibinfo {year}
  {2024})\BibitemShut {NoStop}%
\end{thebibliography}
\end{document}